\documentclass[twocolumn]{revtex4}

\usepackage{xcolor}
\usepackage{amsmath, amsfonts, amssymb, graphicx, color}

\DeclareMathOperator*{\argmin}{arg\,min}

\newcommand{\beq}{\begin{equation}}
\newcommand{\eeq}{\end{equation}}
\newcommand{\<}{\langle}
\renewcommand{\>}{\rangle}
\newcommand{\app}{{Supplementary Text}}

\newcommand{\lastequal}{Corresponding authors. These authors contributed equally.}

\newcommand{\ch}{}

\begin{document}

\newcommand{\deftitle}{{Affinity maturation for an optimal balance between long-term immune coverage and short-term resource constraints}}

\title{\deftitle}

\author{Victor Chard\`es}
\affiliation{Laboratoire de physique de l'\'Ecole normale sup\'erieure,
  CNRS, PSL University, Sorbonne Universit\'e, and Universit\'e de
  Paris, 75005 Paris, France}
\author{Massimo Vergassola}
\affiliation{Laboratoire de physique de l'\'Ecole normale sup\'erieure,
  CNRS, PSL University, Sorbonne Universit\'e, and Universit\'e de
  Paris, 75005 Paris, France}
\author{Aleksandra M. Walczak}
\thanks{\lastequal}
\affiliation{Laboratoire de physique de l'\'Ecole normale sup\'erieure,
  CNRS, PSL University, Sorbonne Universit\'e, and Universit\'e de
  Paris, 75005 Paris, France}
\author{Thierry Mora}
\thanks{\lastequal}
\affiliation{Laboratoire de physique de l'\'Ecole normale sup\'erieure,
  CNRS, PSL University, Sorbonne Universit\'e, and Universit\'e de
  Paris, 75005 Paris, France}

\begin{abstract}
In order to target threatening pathogens, the adaptive immune system performs a continuous reorganization of its lymphocyte repertoire. Following an immune challenge, the B cell repertoire can evolve cells of increased specificity for the encountered strain. This process of affinity maturation generates a memory pool whose diversity and size remain difficult to predict. We assume that the immune system follows a strategy that maximizes the long-term immune coverage and minimizes the short-term metabolic costs associated with affinity maturation. This strategy is defined as an optimal decision process on a finite dimensional phenotypic space, where a pre-existing population of naive cells is sequentially challenged with a neutrally evolving strain. We unveil a trade-off between immune protection against future strains and the necessary reorganization of the repertoire. This plasticity of the repertoire drives the emergence of distinct regimes for the size and diversity of the memory pool, depending on the density of naive cells and on the mutation rate of the strain. The model predicts power-law distributions of clonotype sizes observed in data, and rationalizes antigenic imprinting as a strategy to minimize metabolic costs while keeping good immune protection against future strains.

\end{abstract}

\maketitle

\section{Introduction}


Adaptive immunity relies on populations of lymphocytes expressing diverse antigen-binding receptors on their surface to defend the organism against a wide variety of pathogens. B lymphocytes rely on a two-step process to produce diversity: first a diverse naive pool of cells is generated; upon recognition of a pathogen
the process of affinity maturation allows B cells to adapt their B-cell receptor (BCR) to epitopes of the pathogen  through somatic hypermutation~\cite{Nieuwenhuis1984}. This process, which takes place in germinal centers \cite{Victora2012},
can increase the affinity of naive BCR for the target antigen by up to a thousand fold factor~\cite{Eisen1964}. Through affinity maturation, the immune system generates high-affinity, long-lived plasma cells, providing the organism with humoral immunity to pathogens through the secretion of antibodies---the soluble version of the matured BCR---as well as a pool of memory cells with varying affinity to the antigens~\cite{Weisel2017}. However, the diversity and coverage  
of the memory pool, as well as the biological constraints that control its generation, have not yet been fully explored. 

Analysis of high-throughput BCR sequencing data has revealed long tails in the distribution of clonotype abundances, identifying some very abundant clonotypes as well as many very rare ones~\cite{Weinstein2009,Mora2019c}. Additionally, many receptors have similar sequences and cluster into phylogenetically related lineages~\cite{Kocks1988,Kleinstein2003,Kepler2013,
Yaari2015a,
Ralph2016}. These lineages have been used to locally trace the evolution of antibodies in HIV patients~\cite{Liao2013,Nourmohammad2019} and in influenza vaccinees \cite{Jiang2013,Horns2019}.
Memory B-cell clones are more diverse and less specific to the infecting antigen than antibody-producing plasma cells \cite{Smith1997,Weisel2016}. This suggests that the immune system is trying to anticipate infections by related pathogens or future escape mutants \cite{Viant2020}.

Theoretical approaches have attempted to qualitatively describe affinity maturation as a Darwinian co-evolutionary process, and studied optimal affinity maturation schemes~\cite{Oprea1997,Oprea2000,Kepler1993,Kepler1993}, as well as optimal immunization schedules to stimulate antibodies with large neutralizing capabilities~\cite{Wang2015a, Sachdeva2020, Molari2020}. Most of these approaches have been limited to short timescales, often with the goal of understanding the evolution of broadly neutralizing antibodies. Here we propose a mathematical framework to explore the trade-offs that control how the large diversity of memory cells evolves over a lifetime.  

Despite long-lasting efforts to describe the co-evolution of pathogens and hosts immune systems~\cite{Grenfell2004,Blanquart2013,Cobey2015,Koelle2006,Marchi2021}, and recent theoretical work on optimal schemes for using and storing memory in the presence of evolving pathogens \cite{Schnaack2021a}, few theoretical works have described how the B-cell memory repertoire is modified by successive immunization challenges. Early observations in humans~\cite{Francis1960} have shown that sequential exposure to antigenically drifted influenza strains was more likely to induce an immune response strongly directed towards the first strain the patients were exposed to~\cite{Cobey2017}. This immune imprinting with viral strains encountered early in life was initially called ``original antigenic sin,'' as it can limit the efficiency of vaccination~\cite{Hoskins1979}. This phenomenon has been observed in a variety of animal models and viral strains~\cite{Vatti2017}.  Secondary infections with an antigenically diverged influenza strain can reactivate or ``backboost'' memory cells specific to the primary infecting strain~\cite{Kim2009}. This response is characterized by lower binding affinity but can still have in-vivo efficiency thanks of cross-reactive antibodies~\cite{Linderman2016}. There is a long-standing debate about how detrimental ``original antigenic sin'' is~\cite{Yewdell2020,Worobey2020}. However, the question of under what circumstances an immune response based on memory re-use is favourable has not been addressed. 

We build a theoretical framework of joint virus and repertoire evolution in antigenic space, and investigate how {\ch acute} infections by evolving pathogens have shaped, over evolutionary timescales, the B-cell repertoire response and re-organization. Pathogens causing acute infections {\ch may be encountered multiple times over time scales of years, especially when} they show a seasonal periodicity, while the maturation processes in the B-cell repertoire take place over a few weeks. This observation allows us to consider that affinity maturation happens in a negligible time with respect to the reinfection period. Within this approximation, we investigate the optimal immune maturation strategies using a framework of discrete-time decision process. We show the emergence of three regimes---monoclonal memory response, polyclonal memory response, and naive response---as trade-offs between immune coverage and resource constraint. Additionally, we demonstrate that reactivation of already existing memory clonotypes can lead to self-trapping of the immune repertoire to low reactivity clones, opening the way for ``original antigenic sin.''

\section{Results}


\subsection{Affinity maturation strategies for recurring infections}

\begin{figure}
\begin{center}
\includegraphics{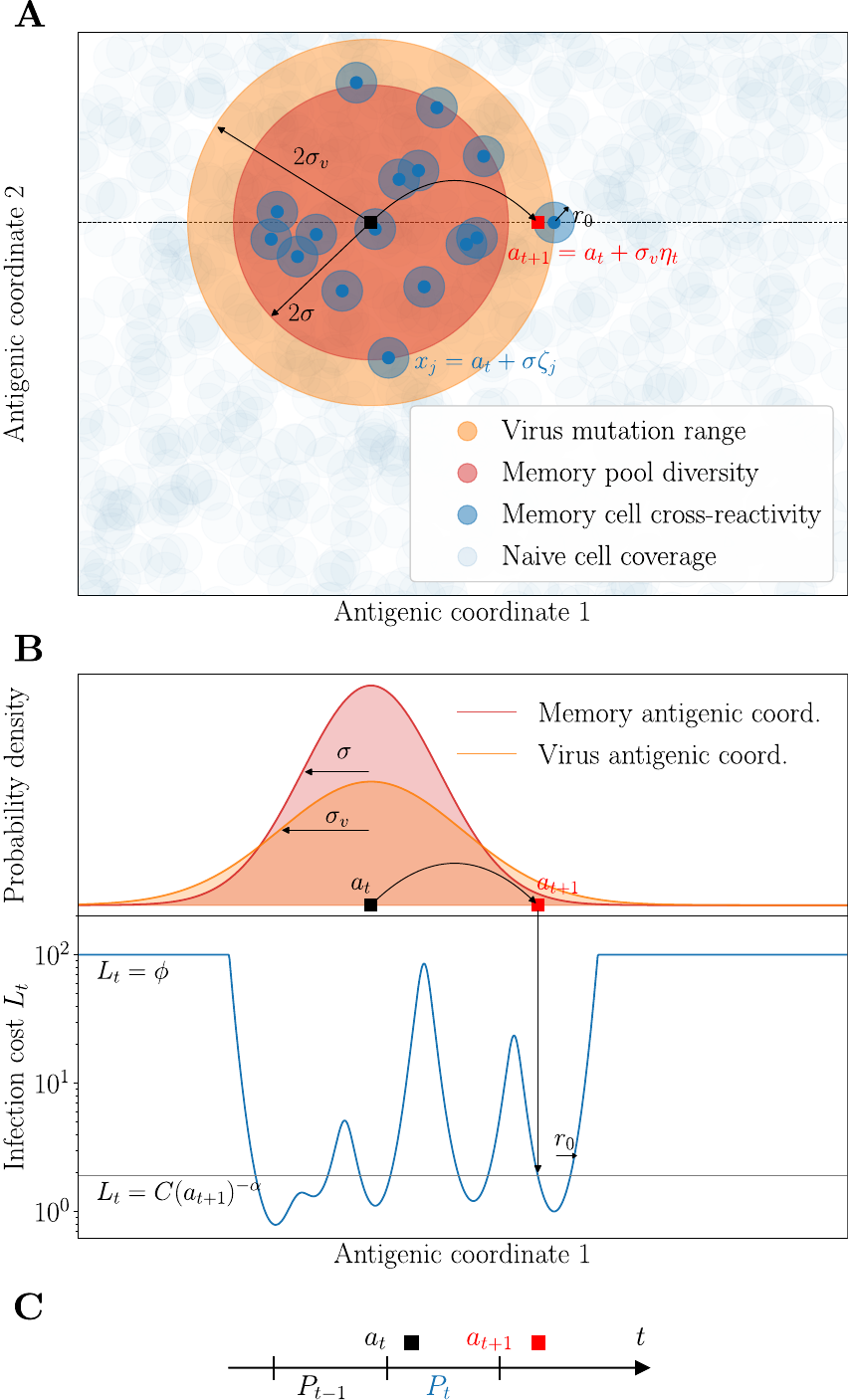}
\caption{{\bf Model of sequential affinity maturation. A.} An infecting strain is defined by its position $a_n$ in antigenic space (dark square). In response, the immune system creates $m$ new memory clonotypes $x_j$ (blue points) from a Gaussian distribution of width $\sigma$ centered in $a_t$ (red area). These new clonotypes create a cost landscape (blue areas) for the next infection, complemented by a uniform background of naive cells (light blue). The next infecting strain (red square) is drawn from a Gaussian distribution of width $\sigma_v$ centered in $a_t$ (orange area). The position of this strain on the infection landscape is shown with the arrow. Antigenic space is shown in 2 dimensions for illustration purposes, but can have more dimensions in the model.
  {\bf B.} Cross-section of the distributions of memories and of the next strain, along with the infection cost landscape $L_t$ (in blue). Memories create valleys in the landscape, on a background of baseline naive protection $\phi$. {\bf C.} Sequential immunization. Strain $a_t$ modifies the memory repertoire into $P_t$, which is used to fight the next infection $a_{t+1}$. $P_t$ is made of all newly created clonotypes (blue points in A) as well as some previously existing ones (not shown). Clonotype abundances are boosted following each infection as a function of the cross-reactivity, and each individual cell survives from one challenge to the other with a probability $\gamma$.}
\label{fig:fig0}  
\end{center}
\end{figure}

B cells recognize pathogens through the binding of their BCR to parts of the pathogen's proteins, called epitopes, {\ch which we refer to as ``antigens'' for simplicity}. To model this complex protein-protein interaction problem, we assume that both receptors and antigens may be projected into an effective, $d$-dimensional antigenic space  (Fig.~\ref{fig:fig0}), following the ``generalized shape space'' idea pioneered by Perelson and Oster \cite{Perelson1979}. Receptor-antigen pairs at close distance in that space bind well, while those that are far away bind poorly. Specifically, we define a cross-reactivity function $0\leq f\leq 1$ quantifying the binding affinity between antigen $a$ and receptor $x$, which we model by a stretched exponential,
{\ch $f(x, a) = e^{-(\|x-a\|/r_0)^q}$}. {\ch This choice of function is the simplest that allows for introducing a cross-reactivity radius, $r_0$, while controlling how sharply recognition is abrogated as the distance between antigen and receptor oversteps that radius, through the stretching exponent $q$.}

For simplicity, we focus on a single pathogen represented by its immunodominant antigen, so that each viral strain is represented by a single point $a_t$ in antigenic space (black square), where $t=1,2,\ldots$ is a discrete time counting the number of re-infections. It is difficult to estimate the rate of re-infections or exposures to the same pathogen. It can be fairly high in humans, where individuals are exposed to the most common viruses from less than once to several times a year \cite{Cohen2021}. The numbers of lifetime exposures would then range from a few to a few hundreds.

The B cell repertoire, on the other hand, is represented by a collection of antigenic coordinates corresponding to each receptor clonotype.
We distinguish memory cells (dark blue circles in Fig.~\ref{fig:fig0}A), denoted by $P_t$, which have emerged in response to the presence of the virus, and a dense background of naive cells $N$ (light blue circles) which together provide a uniform but weakly protective coverage of any viral strain (subsumed into the parameter $\phi$ defined later).

The viral strain evolves randomly in antigenic space, sequentially challenging the existing immune repertoire. This assumption is justified by the fact that for acute infections with a drifting viral strain, such as influenza, the immune pressure exerted on the strain does not happen in hosts but rather at the population level \cite{Grenfell2004}. Viral evolution is not neutral, but it is unpredictable from the point of view of individual immune systems. Specifically, we assume that, upon reinfection, the virus is represented by a new strain, which has moved from the previous antigenic position $a_t$ to the new one $a_{t+1}$ according to a Gaussian distribution with typical jump size $\sigma_v$, called ``divergence" {\ch (see Methods)}.

{\ch Upon infection by a viral strain at $a_{t}$, available cross-reactive memory or naive cells will produce antibodies whose affinities determine the severity of the disease. We quantify the efficiency of this early response to the strain $a_{t}$ with an infection cost $I_t$:
\beq
I_t = \min\Big[\phi, \Big(\sum_{x\in P_{t-1}}n_{x,t}f(x,a_t)\Big)^{-\alpha} \Big],
\eeq
where $\phi>0$ is a maximal cost corresponding to using naive cells, {\ch and where $n_{x,t}$ denotes the size of clonotype $x$ at time $t$}. This infection cost is a decreasing function of the coverage of the virus by the pre-existing memory repertoire, $P_{t-1}$, $C(a_t)=\sum_{x\in P_{t-1}} n_{x,t}f(x,a_t)$, with a power $\alpha$ governing how sharp that dependence is.} Intuitively, the lower the coverage, the longer it will take for memory cells to mount an efficient immune response and clear the virus, incurring larger harm on the organism \cite{Mayer2015,Mayer2019}.

When memory coverage is too low, the {\ch naive B-cell repertoire as well as the rest of the immune system (including its innate branch as well as T-cell cytotoxic activity)} {\ch still offers some protection}, incurring a maximal cost fixed to $\phi$. {\ch Memory cells respond more rapidly than naive cells, which is indirectly encoded in our model by the naive cost $\phi$ being larger than the cost when specific memory cells are present (of order 1 or less).}
In the \app{} we show how this naive cut-off may be derived in a model where the immune system activates its memory and naive compartments in response to a new infection, when naive clonotypes are very numerous but offer weak protection. In that interpretation, $\phi$ scales like the inverse density of naive cells. We will refer to $\phi^{-1}$ as ``naive density,'' {\ch although one should keep in mind that this basal protection levels also includes other arms of the immune system.}
In Fig.~\ref{fig:fig0}B we plot an example of the infection cost along a cross-section of the antigenic space.

{\ch After this early response, activated memory cells proliferate and undergo affinity maturation to create {\it de novo} plasma and memory cells targeting the infecting strain.} To model this immune repertoire re-organization in response to a new infection $a_t$, we postulate that its strategy has been adapted over evolutionary timescales to maximize the speed of immune response to subsequent challenges, given the resource constraints imposed by affinity maturation \cite{Mayer2015}. {\ch This strategy dictates the stochastic rules according to which the BCR repertoire evolves from $P_{t-1}$ to $P_t$ as a result of affinity maturation (Fig.~\ref{fig:fig0}C).}

We consider the following rules inspired by known mechanisms available to the immune system \cite{Victora2012}. {\ch After the infection by $a_t$ has been tackled by existing receptors, and the infection cost has been paid}, new receptors are matured to target {\ch future versions of} the virus. {\ch Their number $m_t$ is distributed according to a Poisson law, whose mean is controlled by the cost of infection, $\bar m (I_t)$}.  {\ch This dependence accounts for the feedback of the early immune response on the outcome of affinity maturation, consistent with extensive experimental evidence of the history dependence of the immune response \cite{Oidtman2021}.}
Each new receptor is roughly located around $a_t$ in antigenic space with some {\ch added noise} $\sigma (I_t)$, {\ch and starts with clonotype size $n_{x,t}=1$ by convention. The diversification parameter $\sigma$ can be tuned by the immune system through the permissiveness of selection in germinal centers, through specific regulation factors induced at the early stage of affinity-based selection \cite{Nakagawa2021}: $\sigma=0$ means that affinity maturation only keeps the best binders to the antigens, while $\sigma>0$ means that selection is weaker.}

{\ch At the same time}, each clonotype $x\in {P}_{t-1}$ from the previous repertoire may be reactivated and be subsequently duplicated through cell divisions \cite{Viant2020}, with probability $\mu f(x,a_t)$ {\ch (see Methods)}, proportional to the cross-reactivity, where $0\leq\mu\leq 1$ is a proliferation parameter. These previously existing cells and their offspring may then die before the next infection. We denote by $\gamma$ their survival probability, so that the average lifetime of each cell is $(1-\gamma)^{-1}$. The proliferation and death parameters $\mu$ and $\gamma$ are assumed to be constrained and fixed. {\ch The net mean growth is thus given by $\<n_{x,t}\>=(1+\mu f(x,a_t))\gamma n_{x,t-1}$. $\Gamma\equiv (1+\mu)\gamma$ is defined as the maximum growth factor.  At the end of the process, the updated repertoire $P_t$ combines the result of this proliferation and death process applied to $P_{t-1}$ with the new receptors obtained from affinity maturation.}

\begin{figure*}[t!]
\begin{center}
\includegraphics{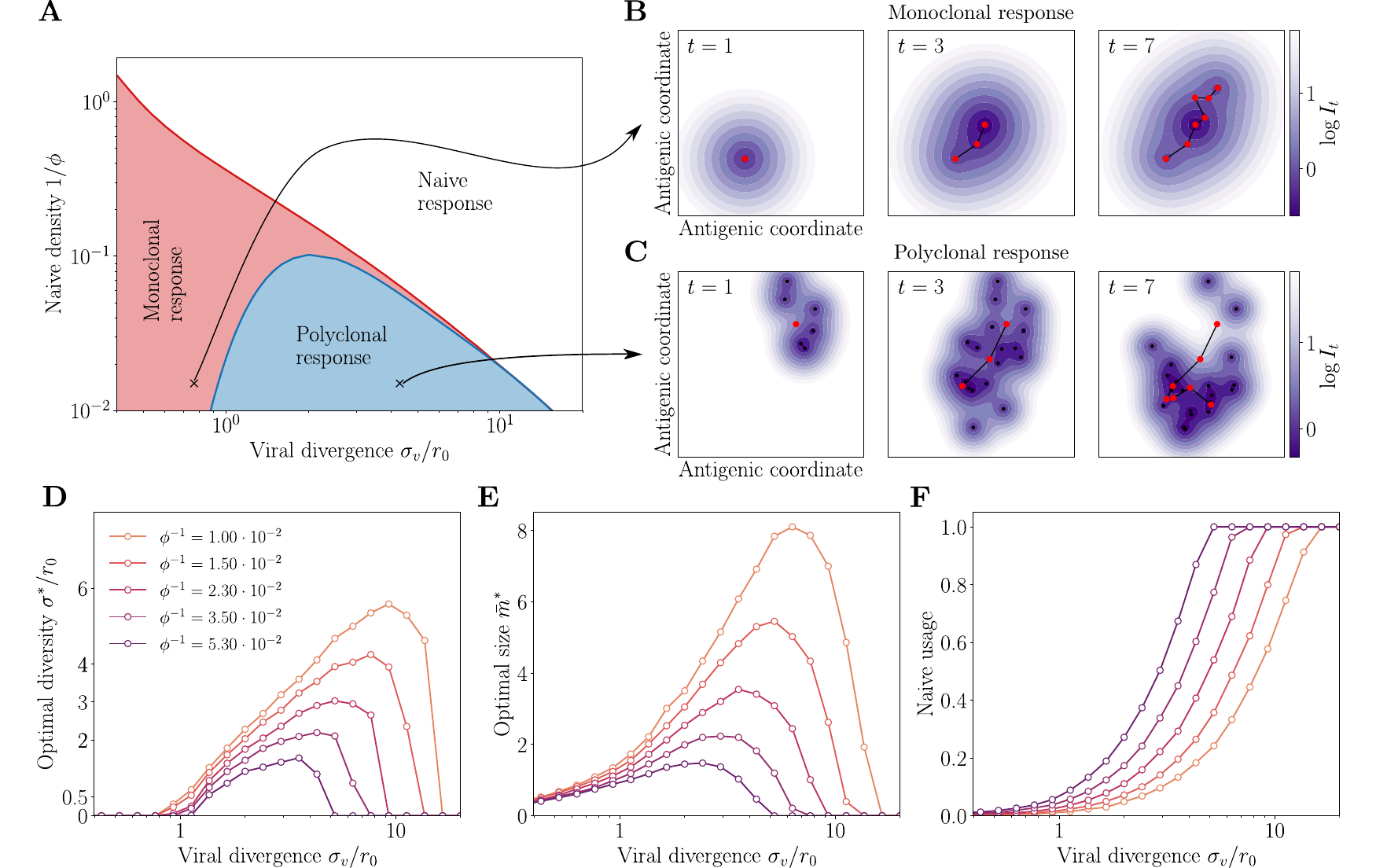}
\caption{{\bf Regimes of affinity maturation. A.} Phase diagram of the model as a function of the naive coverage $1/\phi$, and viral divergence $\sigma_v$, in a two-dimensional antigenic map. Three phases emerge: monoclonal memory (red), polyclonal memory (purple) and naive response (white). {\bf B-C.} Snapshots in antigenic space of the sequential immunization by a viral strain in the {(B)} monoclonal and {(C)} polyclonal phases. We show the viral position (red dots), memory clonotypes (black dots), and viral trajectory (black line). The colormap shows the log infection cost. Parameters $\sigma_v$ and $\phi$ correspond to the crosses on the phase diagram in A, with their respective optimal $\sigma^*$, $\bar m^*$ (see arrows). {\bf D.} Diversity $\sigma^*$, {\bf E.} optimal size $\bar m^*$, and {\bf F.} frequency of naive cell usage in response to an immunization challenge for different naive coverages $1/\phi$. Parameters values: $\kappa = 3.3$, $\alpha = 1$, $q = 2$, $d = 2$,  $\gamma = 0.85$, $\mu = 0.5$.}
\label{fig:fig1}  
\end{center}
\end{figure*}

To assess the performance of a given strategy, we define an overall cost function at each time step:
{\ch
  \beq\label{eq:ln}
L_t=I_t+\kappa m_t.
\eeq
The second term $\kappa m_t$} corresponds to a plasticity cost encoding the resources necessary to generate and maintain new memory clonotypes with affinity maturation. {\ch This cost enforces a minimal homeostatic constraint on the memory repertoire. We neglect any dependence of the cost on the diversification $\sigma$, which is secondary and would require adding additional parameters without affecting the qualitative picture.} We assume that, over evolutionary timescales, the immune system has minimized the average cumulative cost $\<L_t\>$ over a large number of infections {\ch (see Methods)}.
This optimization yields the optimal parameters of the strategy, {\ch namely the best functions $m^*(I)$ and $\sigma^*(I)$ describing the extent and diversity of affinity maturation and how they should depend on the strength of the infection $I$. For the sake of simplicity, in the next three sections we will specialize to the case of constant functions $m(I)\equiv m$ and $\sigma(I)\equiv\sigma$. We will come back to the general case in the last section of the results.}


\subsection{Phase diagram of optimal affinity maturation strategies}
\label{sec:phase_diagram}

We obtain optimal {\ch constant strategies $\bar m(I)=m^*,\sigma(I)=\sigma^*$}, by minimizing the simulated long-term cost 
$\mathcal{L}(\mathcal{S})$ (Eq.~\ref{eq:gen_loss}) in a 2-dimensional antigenic space (see Methods for details of the simulation, optimization procedures, and phase determination). By varying two key parameters, the cost $\phi$ associated to the use of the naive repertoire, and the virus divergence $\sigma_{v}$, we see a phase diagram emerge with three distinct phases: the naive, monoclonal response, and polyclonal response phases (Fig.~\ref{fig:fig1}A). In {Figs.~\ref{fig:fig1}B-C we show examples of the stochastic evolution of memory repertoires with optimal rules in the two  phases (monoclonal and polyclonal responses). Figs.~\ref{fig:fig1}D-F show the behaviour of the optimal parameters, as well as the fraction of infections for which the naive repertoire is used (when the maximal infection cost $\phi$ is paid). The general shape and behaviour of this phase diagram depends only weakly on the parameter choices (see SI Appendix, Fig.~S1).

When the naive repertoire is sufficiently protective (small $\phi$), or when the virus mutates too much between infections (large $\sigma_v$), the optimal strategy is to produce no memory cells at all ($\bar m^*=0$), and rely entirely on the naive repertoire, always paying a fixed cost $L_t=\mathcal{L}=\phi$ (naive phase).

When the virus divergence $\sigma_v$ is small relative to the cross-reactivity range $r_0$, it is beneficial to create memory clonotypes ($\bar m^*>0$), but with no diversity, $\sigma^*=0$ (monoclonal response). In this case, all newly created clonotypes are invested into a single antigenic position $a_t$ that perfectly recognizes the virus. This strategy is optimal because subsequent infections, typically caused by similar viral strains of the virus, are well recognized by these memory clonotypes. 

For larger but still moderate virus divergences $\sigma_v$, this perfectly adapted memory is not sufficient to protect from mutated strains: the optimal strategy is rather to generate a polyclonal memory response, with $\bar m ^{*} > 0$, $\sigma^{*} > 0$. In this strategy, the immune system hedges its bet against future infections by creating a large diversity of clonotypes that cover the vicinity of the encountered strain. {\ch The created memories are thus less efficient against the {\em current} infection, which they never will have to deal with.
The advantage of this strategy is to anticipate future antigenic mutations of the virus.}
This diversified pool of cells with moderate affinity is in agreement with recent experimental observations~\cite{Victora2012,Tas2016, Kuraoka2016,Viant2020}. The diversity of the memory pool is supported by a large number of clonotypes $\bar m^*$ (Fig.~\ref{fig:fig1}D). As the virus divergence $\sigma_v$ is increased, the optimal strategy is to scatter memory cells further away from the encountered strain (increasing $\sigma^*$, Fig.~\ref{fig:fig1}E).  
However, when $\sigma_v$ it is too large, both drop to zero as the naive repertoire takes over (Fig.~\ref{fig:fig1}F). Increasing the naive density $\phi^{-1}$ also favors the naive phase. When there is no proliferation on average, i.e. $\Gamma= (1+\mu)\gamma<1$, there even exists a threshold $\phi_c^{-1}$ above which the naive strategy is always best (SI Appendix Fig.~S1, and Text for estimates of that threshold).

{\ch We derived analytical results and scaling laws in a simplified version of the model, where cross-reactivity is ideally sharp, and where memory cells survive only until the next infection. The results are derived in \app{} and illustrated in SI Appendix, Fig.~S2. In this simplified setting, the monoclonal to polycolonal transition occurs at $\sigma_v\sim r_0$, consistent with the intuition that diversification occurs when the virus is expected to escaped out of the cross-reactivity radius. The polyclonal-to-naive transition occurs at $(\sigma_v/r_0)^d\sim \phi/\kappa$, when basal coverage by the naive repertoire (and rest of the immune system) outcompetes the protection afforded by memory cells relative to their cost $\kappa$.}

In summary, the model predicts the two expected regimes of naive and memory use depending on the parameters that set the costs of infections and memory formation. But in addition, it shows a third phase of polyclonal response, where affinity maturation acts as an anticipation mechanism whose role is to generate a large diversity of cells able to respond to future challenges. The prediction of a less focused and thus weaker memory pool observed experimentally is thus rationalized as a result of a bet-hedging strategy.

\begin{figure*}[t]
\begin{center}
\includegraphics{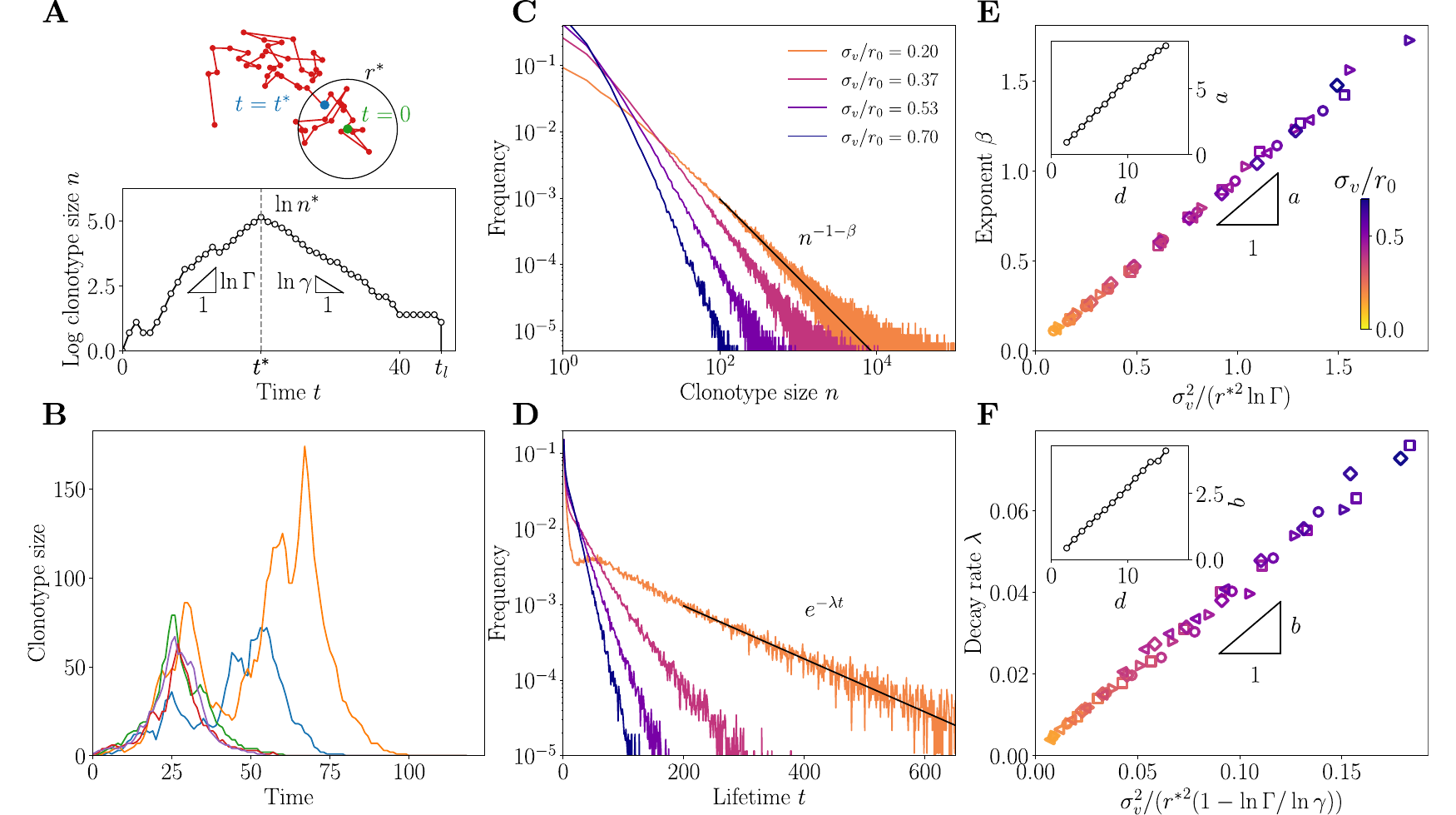}
\caption{{\bf Clonotype dynamics and distribution. A.} Sketch of a recall response generated by sequential immunization with a drifting strain. Clonotypes first grow with multiplicative rate $\Gamma=\gamma(1+\mu)$, until they reach the effective cross-reactivity radius $r^*$, culminating at $n^*$, after which they decay with rate $\gamma$ until extinction at time $t_l$. {\bf B.}  Sample trajectories of clonotypes generated by sequential immunization with a strain of mutability $\sigma_v/r_0 = 0.53$. {\bf C.} Distribution of clonotype size for varying virus mutability $\sigma_v/r_0$. {\bf D.} Distribution of the lifetime of a clonotype for varying virus mutability $\sigma_v / r_0$. From {\bf B} to {\bf D} the proliferation parameters are set to $\gamma = 0.85$, $\gamma = 0.5$ ie. $\Gamma = 1.275$. {\bf E.} Scaling relation of the power law exponent for varying values of the parameters. Inset: dependence of the proportionality factor $a$ on dimension. {\bf F.} Scaling relation of the decay rate $\lambda$ for varying $\sigma_v / r_0$, with scaling of the proportionality factor $b$. In both {\bf{E-F.}}, the different parameters used are $(\gamma = 0.82, \mu = 0.65)$  ie. $\Gamma = 1.353$ (diamonds), $(\gamma = 0.8, \mu = 0.62)$ ie. $\Gamma = 1.296$ (squares), $(\gamma = 0.85, \mu = 0.5)$ ie. $\Gamma = 1.275$ (circles), $(\gamma = 0.87, \mu = 0.4)$ ie. $\Gamma = 1.21$ (triangles $>$), $(\gamma = 0.9, \mu = 0.35)$ ie. $\Gamma = 1.21$ (triangles $<$). From {\bf B} to {\bf F}, the strategy was optimized for $\phi = 100$ and $\kappa = 0.5/(1 - \gamma)$. The color code for $\sigma_v/r_0$ is consistent across the panels {\bf C} to {\bf F}. In this panel, the other parameters used are $\alpha = 1$, $q = 2$, $d = 2$.}
\label{fig:fig3}  
\end{center}
\end{figure*}

\subsection{Population dynamics of optimized immune systems}

We now turn to the population dynamics of the memory repertoire. When the virus drifts slowly in antigenic space (small $\sigma_v$), the same clonotypes get reactivated multiple times, causing their proliferation, provided that $\Gamma=\gamma(1+\mu)>0$. This reactivation continues until the virus leaves the cross-reactivity range of the original clonotype, at which point the memory clone decays and eventually goes extinct (Fig.~\ref{fig:fig3}A). Typical clonotype size trajectories from the model are shown in Fig.~\ref{fig:fig3}B. They show large variations in both their maximal size and lifetime.
The distribution of clonotype abundances, obtained from a large number of simulations, is indeed very broad, with a power-law tail (Fig.~\ref{fig:fig3}C). The lifetime of clonotypes, defined as the time from emergence to extinction, is distributed according to an exponential distribution (Fig.~\ref{fig:fig3}D). The exponents governing the tails of these distributions, $\beta$ and $\gamma$, depend on the model parameters, in particular the divergence $\sigma_v$.

{\ch We can understand the emergence of these distributions using a simple scaling argument, detailed in \app{}. The peak size of a clonotype depends on the number of successive infections by viral strains remaining within a distance $r^*=r_0\ln[\gamma\mu/(1-\gamma)]^{1/q}$ from the clonotype, under which it continues expanding. This number has a long exponential tail with characteristic time $t_s\sim (r^*/\sigma_v)^2$. One can show that this translates into a power law tail for the distribution of clonotype sizes:
  \beq\label{eq:beta}
  p(n^*) \sim \frac{1}{{n^*}^{1+\beta}},\quad\textrm{with } \beta\sim \frac{\sigma_v^2}{{r^*}^2 \ln \Gamma},
  \eeq
  and an exponential tail for the lifetime of clonotypes:
  \beq\label{eq:lambda}
p(t_l)\sim e^{-\lambda t_l},\quad \lambda\sim\frac{\sigma_v^2}{{r^*}^2}{\left(1+\frac{\ln\Gamma}{\ln(1/\gamma)}\right)}^{-1}.
\eeq
This simple scaling argument predicts the exponents $\beta$ and $\lambda$ fairly well: Figs.~\ref{fig:fig3}E-F confirm the validity of the scaling relations \eqref{eq:beta}-\eqref{eq:lambda} against direct evaluation from simulations, for $d=2$ and $q=2$. These scalings still hold for different parameter choices (see SI Appendix, Fig.~S3).
}

These scaling relations are valid up to a geometry-dependent prefactor, which is governed by dimensionality and the shape of the cross-reactivity kernel. In the \app{}, we calculate this prefactor in the special case of an all-or-nothing cross-reactivity function, $q=\infty$.
Generally, $\beta$ increases with $d$, as shown in the insets of Figs.~\ref{fig:fig3}E-F for $q=2$. In higher dimensions, there are more routes to escape the cross-reactivity range, and thus a faster decaying tail of large clonotypes. This effect cannot be explained by having more dimensions in which to mutate, since the antigenic variance is distributed across each dimension, according to $\sigma_v^2/d$. Rather, it results from the absence of antigenic back-mutations: in high dimensions, each mutation drifts away from the original strain with a low probability of return, making it easier for the virus to escape, and rarer for memory clonotypes to be recalled upon infections by mutant strains.

\begin{figure}[t!]
\begin{center}
\includegraphics{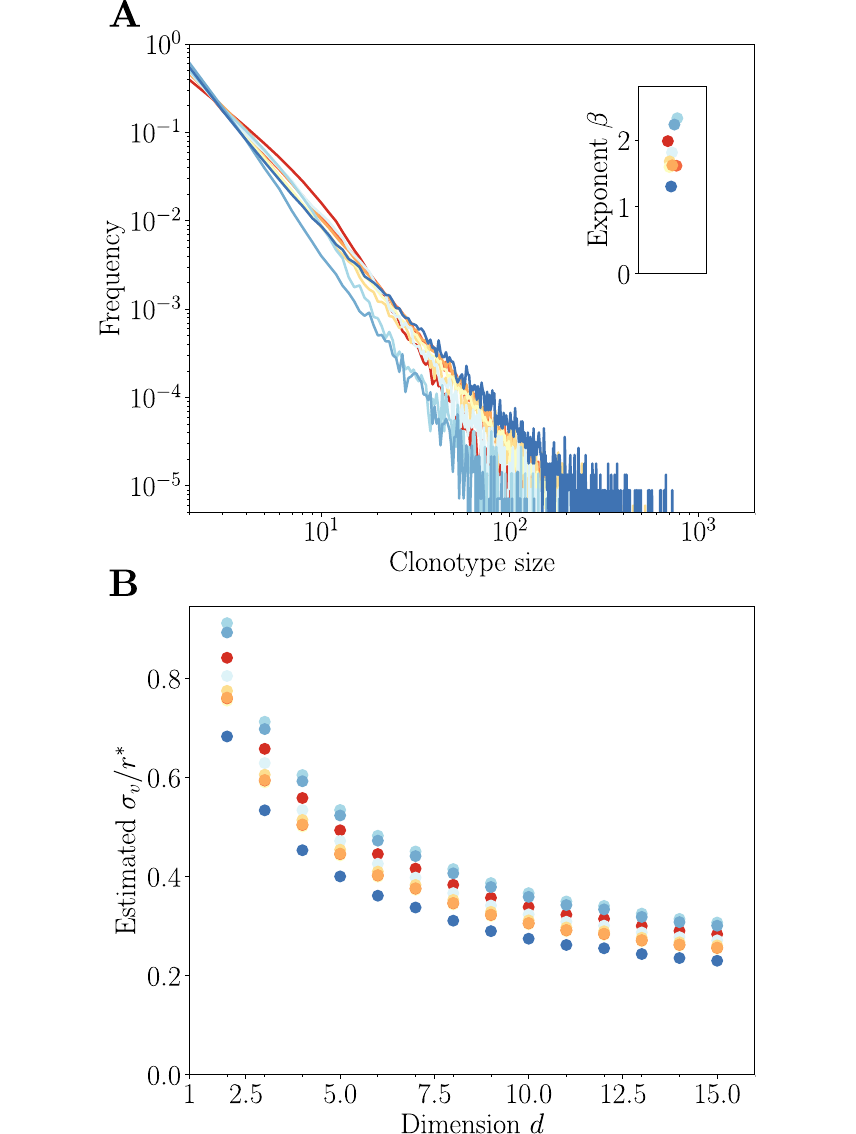}
\caption{{\bf Comparison to repertoire data. A.} Clonotype abundance distribution of IgG repertoires of healthy donors from \cite{Briney2019}. {\bf B.} Estimated mutatibility $\sigma_v$ in units of the rescaled cross-reactivity $r^*$, defined as the antigenic distance at which clonotypes stop growing. $\sigma_v$ is obtained as a function of $d$ by inverting the linear relationship estimated in the inset of Fig.~\ref{fig:fig3}E, assuming $q=2$ and $\Gamma=1.4$ (estimated from \cite{Viant2020}).}
\label{fig:fig4}  
\end{center}
\end{figure}

\subsection{Comparison to experimental clone-size distributions}

The power-law behaviour of the clone-size distribution predicted from the model (Fig.~\ref{fig:fig3}E) {\ch can be directly compared to existing data on bulk repertoires. While the model makes a prediction for subsets of the repertoire specific to a particular family of pathogens, the same power-law prediction is still valid for the entire repertoire, which is a mixture of such sub-repertoires.}
Power laws have been widely observed in immune repertoires: from early studies of repertoire sequencing data of BCR in zebrafish \cite{Weinstein2009,Mora2010a}, to the distribution of clonal family sizes of human IgG BCR \cite{Jiang2013,Spisak2020}, as well as in T-cell receptor repertoires \cite{Mora2019c}. However, these power laws have not yet been reported in the clonotype abundance distribution of human BCR.

To fill this gap, we used publicly available IgG repertoire data of 9 human donors from a recent ultra-deep repertoire profiling study of immunoglobulin heavy-chains (IGH) \cite{Briney2019}. The data was downloaded from Sequence Read Archive and processed as in \cite{Spisak2020}. Repertoires were obtained from the sequencing of IGH mRNA molecules to which unique molecular identifiers (UMI) were appended. For each IGH nucleotide clonotype found in the dataset, we counted the number of distinct UMI attached to it, as a proxy for the overall abundance of that clonotype in the expressed repertoire. The distributions of these abundances are shown for all 9 donors in Fig.~\ref{fig:fig4}A. In agreement with the theory, they display a clear power-law decay $p(n)\sim n^{-1-\beta}$, with $\beta=1.2$-$2.4$.

Since the experimental distribution is derived from small subsamples of the blood repertoire, the absolute abundances cannot be directly compared to those of the model. In particular, subsampling means that the experimental distribution focuses on the very largest clonotypes. Thus, comparisons between model and data should be restricted to the tail behavior of the distribution, namely on its power-law exponent $\beta$. {\ch The bulk repertoire is a mixture of antigen-specific sub-repertoires, each predicted to be a power law with a potentially different exponent. The resulting distribution is still a power-law dominated by the largest exponent.}

We used this comparison to predict from the exponent $\beta$ the virus divergence between infections. To do so, we fit a linear relationship to the inset of Fig.~\ref{fig:fig3}E, and invert it for various values of the dimension $d$ to obtain $\sigma_v/r^*$. We fixed $\Gamma=1.4$, which corresponds to a 40\% boost of memory B cells upon secondary infection, inferred from a 4-fold boost following 4 sequential immunizations reported in mice \cite{Viant2020}. The result is robust to the choice of donor, but decreases substantially with dimension, because higher dimensions mean faster escape, and thus a lower divergence for a given measured exponent (Fig.~\ref{fig:fig4}B). The inferred divergence $\sigma_v$ is always lower than, but of the same order as, the effective cross-reactivity range $r^*$, suggesting that the operating point of the immune system falls in the transition region  between the monoclonal and polyclonal response phases (Fig.~\ref{fig:fig1}A). 

\begin{figure*}[t!]
\begin{center}
\includegraphics{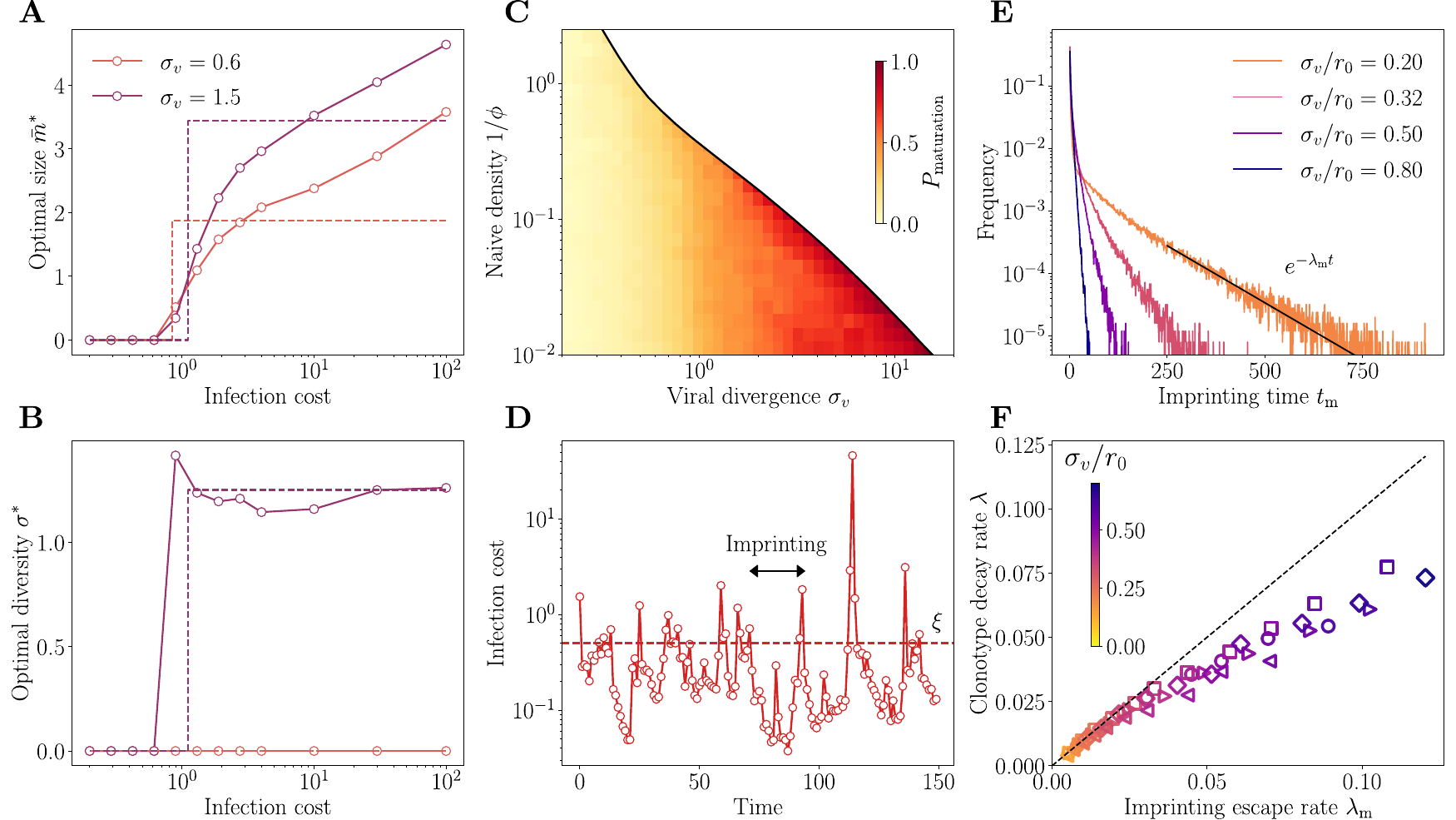}
\caption{{\bf Imprinting and backboosting.} {\ch {\bf A,B.} Optimal regulatory functions for ({\bf A}) the number $ m^*(I)$ and ({\bf B}) the diversity $\sigma^*(I)$ of new memories as a function of the infection cost $I$, for two values for the viral divergence. These functions show a sharp transition from no to some memory formation, suggesting to replace them by simpler step functions (dashed lines). This step function approximation is used in the next panels.}
  {\bf C.}  Frequency of infections
leading to affinity maturation in the optimal strategy. The frequency
increases with the virus divergence $\sigma_v$, up to the point where
the transitions to the naive phase where memory is not used at
all. {\bf D.} Typical trajectory of infection cost in sequential
infections at $\sigma_v/r_0 = 0.5$. When the cost goes beyond the threshold $\xi$, affinity maturation is activated, leading to a drop in infection cost. These periods of sub-optimal memory
describe an ``original antigenic sin,'' whereby the immune system is
frozen in the state imprinted by the last maturation event. {\bf E.} Distribution of imprinting times,
i.e. the number of infections between affinity maturation
events, decays exponentially {\ch with rate $\lambda_m$}. The proliferation parameters in {\bf A} to {\bf E} are set to $\gamma = 0.85$ and $\mu = 0.5$. {\bf  F.} Predicted scaling {\ch of $\lambda_m$ with the clonotype decay rate $\lambda$ from Figs.~\ref{fig:fig3}D and F}. In {\bf{F}}, the different parameters used are $(\gamma = 0.82, \mu = 0.65)$  ie. $\Gamma = 1.353$ (diamonds), $(\gamma = 0.8, \mu = 0.62)$ ie. $\Gamma = 1.296$ (squares), $(\gamma = 0.85, \mu = 0.5)$ ie. $\Gamma = 1.275$ (circles), $(\gamma = 0.87, \mu = 0.4)$ ie. $\Gamma = 1.21$ (triangles $>$), $(\gamma = 0.9, \mu = 0.35)$ ie. $\Gamma = 1.21$ (triangles $<$). The color code for $\sigma_v/r_0$ is consistent across the panels {\bf E} and {\bf F}. From {\bf D} to {\bf F}, the strategy was optimized for $\phi = 100$ and $\kappa = 0.5/(1 - \gamma)$. In this panel, the other parameters used are $\alpha = 1$, $q = 2$, $d = 2$.}
\label{fig:fig5}  
\end{center}
\end{figure*}

\subsection{Inhibition of affinity maturation and antigenic imprinting}
\label{sec:affinity_maturation}

{\ch In this section we come back to general strategies where the process of affinity maturation depends on the immune history through the infection cost $I$ experienced by the system during the early immune response, which controls the number and diversity of newly created memories following that response: $\bar m(I)$, $\sigma(I)$.  The optimization of the loss function \eqref{eq:ln} is now carried out with respect to two {\em functions} of $I$. To achieve this task, we optimize with respect to discretized functions $(\bar m_1, ..., \bar m_n)$  and $(\sigma_1, ..., \sigma_n)$ taken at $n$ values of the infection cost $I$ between 0 and $\phi$. From this optimization, a clear transition emerges between a regime of complete inhibition of affinity maturation ($\bar m^*(I) = 0$) at small infection costs, and a regime of affinity maturation ($\bar m^*(I)>0$) at larger infection costs (Fig.~\ref{fig:fig5}A).  In the phase where affinity maturation occurs, the optimal diversity $\sigma^*(I)$ is roughly constant  (Fig.~\ref{fig:fig5}B).

This transition means that when pre-existing protection is good enough, the optimal strategy is not to initiate affinity maturation at all, to save the metabolic cost $\kappa m_t$.
 This inhibition of affinity maturation is called ``antigenic imprinting,'' and is linked to the notion of ``original antigenic sin,'' whereby the history of past infections determines the process of memory formation, usually by suppressing it. This phenomenon leads to the paradoxical prediction that a better experienced immune system is less likely to form efficient memory upon new infections. Importantly, in our model this behaviour does not stem from a mechanistic explanation, such as competition for antigen or T-cell help between the early memory response and germinal centers, but rather as a long-term optimal strategy maximizing immune coverage while minimizing the costs of repertoire re-arrangement.

To simplify the investigation of antigenic imprinting, we approximate the optimal strategies in Fig.~\ref{fig:fig5}A and B by step functions, with a suppressed phase, $\bar m(I) = 0$, for $I < \xi$, and an active phase, $\bar m(I)\equiv \bar m > 0$ and $\sigma(I)\equiv \sigma$, for $I > \xi$. The threshold $\xi$ is left as an optimization parameter, in addition to $\sigma$ and $\bar m$.}
Optimizing with respect to these three parameters, we observe that the frequency of affinity maturation events mostly depends on $\sigma_v$ (Fig.~\ref{fig:fig5}C). While this threshold remains approximately constant, the frequency of affinity maturation events increases as  $\sigma_v$ increases. At small $\sigma_v$, the optimal strategy is to extensively backboost existing memory cells; for large $\sigma_v$, the growing unpredictability of the next viral move makes it more likely to have recourse to affinity maturation.
In other words, when the virus is stable (low $\sigma_v$), the immune system is more likely to capitalize on existing clonotypes, and not implement affinity maturation, because savings made on the plasticity cost outweigh the higher infection cost. As the virus drifts away with time, this infection cost also increases, until it reaches the point where affinity maturation becomes worthwile again.

Trajectories of the infection cost show the typical dynamics induced by backboosting, with long episodes where existing memory remains sufficient to keep the cost below $\xi$ (Fig.~\ref{fig:fig5}D), interrupted by infections that fall too far away from existing memory, triggering a new episode of affinity maturation and concomitant drop in the infection cost.

We call the time between affinity maturation events $t_{\mathrm{m}}$. Its mean $\<t_{\mathrm{m}}\>$ is equal to the inverse of the frequency of maturation events, and thus decreases with $\sigma_v$. Its distribution, shown Fig.~\ref{fig:fig5}E, has an exponential tail with exponent $\lambda_{\mathrm{m}}$. 
The exponential tail of the distribution of $t_{\mathrm{m}}$ is dominated by episodes where the viral strain drifted less than expected. In that case, the originally matured clonotype grows to a large size, offering protection for a long time, even after it has stopped growing and only decays. {\ch We therefore expect that in the case of a slowly evolving virus $\sigma_v \ll r_0$, the escape rate from the suppressed phase is given by the clonotype decay rate: $\lambda_{\mathrm{m}} \sim \lambda$. We verify this prediction in Fig.~\ref{fig:fig5}F. Interestingly, for slowly evolving viruses, the typical clonotype lifetime diverges, leading to a lifelong imprinting by the primary immune challenge. Conversely, as the viral divergence $\sigma_v$ grows, the imprinting time decays faster than the typical clonotype lifetime and the extent of the imprinting phenomenon is limited.}

\section{Discussion}

Adaptive immunity coordinates multiple components and cell types across entire organisms over many space and time scales. Because it is difficult to empirically characterize the immune system at the system scale, normative theories have been useful to generate hypotheses and fill the gap between observations at the molecular, cellular, and organismal scales \cite{Chakraborty2017,Altan-Bonnet2020}. Such approaches include clonal selection theory \cite{Burnet1957}, or early arguments about the optimal size and organization of immune repertoires \cite{Perelson1979,Perelson1976,Perelson1978}, and of affinity maturation \cite{Kepler1993,Oprea2000}. While these theories do not rely on describing particular mechanisms of immune function, they may still offer quantitative insights and help predict new mechanisms or global rules of operation.

Previous work developed models of repertoire organization as a constrained optimization problem where the expected future harm of infection, or an {\em ad hoc} utility function, is minimized~\cite{Mayer2015, Mayer2019, Marsland2021,Schnaack2021a}. In Ref.~\cite{Mayer2019}, it was assumed clonotypes specific to all antigens are present at all times in the repertoire; the mechanism of immune memory then merely consists of expanding specific clonotypes at the expense of others. This assumption describes T-cell repertoires well, where there are naive cells with good affinity to essentially any antigen \cite{Moon2007}. For B cells the situation is more complex because of affinity maturation. In addition, re-organizing the repertoire through mutation and selection has a cost, and is subject to metabolic and physical constraints.

Our work addresses these challenges by proposing a framework of discrete-time decision process to describe the optimal remodeling of the B-cell repertoire following immunization, through a combination of affinity maturation and backboosting. While similar to \cite{Schnaack2021a}, our approach retains the minimal amount of mechanistic details and focuses on questions of repertoire remodeling, dynamics, and structure. {\ch The specific choices of the cost functions were driven by simplicity, while still retaining the ability to display emergent behaviour. Generalizing the metabolic cost function to include e.g. costs of diversification (through a dependence on $\sigma$) or of cell proliferation is not expected to affect our results qualitatively, although it may shift the exact positions of the transition boundaries.}

We investigated strategies that maximize long-term protection against subsequent challenges and minimize short-term resource costs due to the affinity maturation processes. Using this model, we observed that optimal strategies may be organized into three main phases as the pathogen divergence and naive coverage are varied.
We expect these distinct phases to co-exist in the same immune system, as there exists a wide range of pathogen divergences, depending on their evolutionary speed and typical frequency of recurrence.

For fast recurring or slowly evolving pathogens, the monoclonal response ensures a very specific and targeted memory. This role could be played by long-lived plasma cells. These cells are selected through the last rounds of affinity maturation, meaning that they are very specific to the infecting strain \cite{Akkaya2020}. Yet, despite not being called memory cells, they survive for long times in the bone marrow, providing long-term immunity.

For slow recurring or fast evolving pathogens, the polyclonal response provides a diverse memory to preempt possible antigenic drift of the pathogen. The function could be fulfilled by memory B cells, which are released earlier in the affinity maturation process, implying that they are less specific to the infecting strain, but collectively can cover more immune escape mutations. {\ch While affinity maturation may start from both memory or naive B cells during sequential challenges, the relative importance of each is still debated \cite{Turner2020,Mesin2020,Wong2020b}}. Our model does not commit {\ch on this question} since we assume that the main benefit of memory is on the infection cost, rather than its re-use in subsequent rounds of affinity maturation.

{\ch For simplicity our model assumed random evolution of the virus. However, there is evidence, backed by theoretical arguments, that successive viral strains move in a predominant direction in antigenic space, as a result of immune pressure by the host population \cite{Bedford2012a,Bedford2014,Marchi2021}. While it is unlikely that the immune system has evolved to learn how to exploit this persistence of antigenic motion in a specific manner, such a bias in the random walk is expected to affect the optimal strategy, as we checked in simulations (Fig.~S4). The bias of the motion effectively increases the effective divergence of the virus, favoring the need for more numerous and more diverse memory cells. However, it does not seem to affect the location of the polyclonal-to-naive transition.}

{\ch The model is focused on acute infections, motivated by the assumption that recurring infections and antigenic drift are the main drivers of affinity maturation evolution. However, much of the model and its results can be re-interpreted for chronic infections. In that context, the sequential challenges of our model would correspond to selective sweeps in the viral population giving rise to new dominant variants. While the separation of time scales between the immune response and the rate of reinfections would no longer hold, we expect some predictions, such as the distribution of clonotype sizes and the emergence of imprinting, to hold true. Chronic infections also imply that the virus evolves as a function of how the immune system responds. Including this feedback would require a game-theoretic treatment. We speculate that it would drive antigenic motion in a persistent direction, as argued earlier and evaluated in Fig.~S4.}

{\ch We investigated  strategies where the outcome of affinity maturation is impacted by the efficiency of the early immune response. It is known that the extrafollicular response can drastically limit antigen availability and T-cell help, decreasing the extent of affinity maturation and the production of new plasma and memory cells \cite{Arulraj2021}. Our general framework allows for but does not presume the existence of such negative feedbacks. Instead, they naturally emerge from our optimization principle. We further predict a sharp transition from no to some affinity maturation as a function of the infection cost. This prediction can be interpreted as the phenomenon of antigenic imprinting widely described in sequential immunization assays~\cite{Kim2009}, or ``original antigenic sin'' \cite{Vatti2017}.  
It implies that having been exposed to previous strains of the virus is detrimental to mounting the best possible immune response. 
Importantly, while antigenic imprinting has been widely described in the literature, no evolutionary justification was ever provided for its existence. Our model explains it as a long term optimal strategy for the immune system, maximizing immune coverage while minimizing repertoire re-arrangements (encoded in the cost $\kappa m$).}

{\ch We believe this framework can be generalized to investigate interactions between slow and fast varying epitopes, which are known to be at the core of the low effectiveness of influenza vaccines \cite{Cobey2017}.} When during sequential challenges only one of multiple epitopes changes at a time, it may be optimal for the immune system to rely on its protections against the invariant epitopes. Only after all epitopes have escaped immunity does affinity maturation get re-activated concomitantly to a spike of infection harm, similar to our result for a single antigen.

{\ch Our model can explain previously reported power laws in the distributions of abundances of B-cell receptor clonotypes. However, there exist alternative explanations to such power laws \cite{Desponds2016,Gaimann2020} that do not require antigenically drifting antigens.
Our model predictions could be further tested in a mouse model, by measuring the B-cell recall response to successive challenges \cite{Kim2009}, but with epitopes carefully designed to drift in a controlled manner, to check the transition predicted in Fig.~\ref{fig:fig5}A. While not directly included in our model, our result also suggest that the size of the inoculum, which would affect the infection cost, should also affect backboosting. This effect could also be tested in mouse experiments.
The predicted relationships between viral divergence and the exponents of the power law and clonotype lifetimes (Figs.~\ref{fig:fig3}E and F) could be tested in longitudinal human samples, by sequencing sub-repertoires specific to pathogens with different rates of antigenic evolution. This would require to computationally predict what B-cell receptors are specific to what pathogen, which in general is difficult. 
}

We only considered a single pathogenic species at a time, with the assumption that pathogens are antigenically independent, so that the costs relative to different pathogens are additive. Possible generalizations could include secondary infections, as well as antigenically related pathogens showing patterns of cross-immunity (such as cowpox and smallpox, or different groups of influenza), which could help us shed light on complex immune interactions between diseases and serotypes, such as negative interference between different serotypes of the Dengue fever leading to hemorrhagic fever, or of the human Bocavirus affecting different body sites or \cite{Vatti2017}.

\section{Materials and Methods}

{\ch
\subsection{Mathematical model}
The viral strain is modeled by its antigenic position, which follows a discrete random walk:
\beq
a_{t+1}= a_t+\sigma_v \eta_{t+1},
\eeq
where $\eta_t$ is a normally distributed $d$-dimensional variable with $\<\eta_t\>=0$ and $\<\eta_{t}\cdot\eta_{t'}\>=\delta_{tt'}$.

The positions of newly created memory receptors are drawn at random according to: $x_j=a_t+\sigma \xi_j$, $j=1,\ldots,m_t$, where $\xi_j$ is normally distributed with $\<\xi_j\>=0$ and $\<\xi_j^2\>=1$. Their initial sizes are set to $n_{x_j,t}=1$.

Upon further stimulation, the new size $n'_{x,t}$ of a pre-existing clonotype right after proliferation is given by $n_{x,t}'-n_{x,t-1}\sim \mathrm{Binom}(n_{x,t-1},\mu f(x,a_t))$, where $f(x,a)=e^{-(\Vert x-a\Vert/r_0)^q}$ is the cross-reactivity Kernel. After proliferation, each memory may die with probability $\gamma$, so that the final clonotype size after an infection cycle is given by $n_{x,t}\sim\mathrm{Binom}(n'_{x,t},\gamma)$.

The objective to be minimized is formally defined as a long-term average:
\beq\label{eq:gen_loss}
\mathcal{L}(m,\sigma)=\lim_{T\to\infty}\frac{1}{T}\sum_{t=1}^T L_t.
\eeq
The optimal strategy is defined as:
\begin{equation}
  (\bar m^*,\sigma^*)=\argmin_{(m,\sigma)}\mathcal{L}(m,\sigma),
\end{equation}
where $m^*$ and $\sigma^*$ are, in the general case, full functions of the infection cost $I$, $m^*(I)$ and $\sigma^*(I)$. In all results except in the last section of the Results, we use the Ansatz of constant functions, $m(I)\equiv m$, $\sigma(I)\equiv\sigma$. In the last section of the Results, we first perform optimization over discretized functions $m=(m_1,\ldots,m_n)$, $\sigma=(\sigma_1,\ldots,\sigma_n)$ defined over $n$ chosen values of $I=(I_1,\ldots,I_n)$. Then, we parametrize the functions as step functions: $\sigma(I)=0$ and $m(I)=0$ for $I<\xi$, and $\sigma(I)=\sigma$ and $m(I)=m$ for $I>\xi$, and optimize over the 3 parameters $\sigma,m,\xi$.
}

\subsection{Monte-Carlo estimation of the optimal strategies}
The average cumulative cost $\mathcal{L}$ in Eq.~\ref{eq:gen_loss} is approximated by a Monte-Carlo method. To ensure the simulated repertoire reaches stationarity, we start from a naive repertoire and discard an arbitrary number of initial viral encounters. Because the process is ergodic, simulating a viral-immune trajectory over a long time horizon is equivalent to simulating $M$ independent trajectories of smaller length $T$. To ensure the independence of our random realizations across our $M$ parallel streams we use the random number generators method \textit{split} provided in Tina's RNG library \cite{Bauke2007}. The cumulative cost function $\mathcal{L}$ is convex for the range of parameters tested. To optimize $\mathcal{L}$ under positivity constraints for the objective variables $\sigma$, $\bar m$ and $\xi$, we use Py-BOBYQA \cite{Cartis2019}, a coordinate descent algorithm supporting noisy objective functions.

The polyclonal to monoclonal (red curve) and memory to naive (blue curve) boundaries of the phase diagrams in SI Appendix, Fig.~S2 and Fig.~S3 are obtained by respectively solving $\partial \mathcal{L}/\partial \sigma = 0$ in the monoclonal phase and $\partial \mathcal{L}/\partial \bar m = 0$ in the naive phase. Both these derivatives can be approximated by finite differences with arbitrary tolerances on $\sigma$ and $\bar m$. We fix the tolerance on $\sigma$ to 0.2 and the tolerance on $\bar m$ to 0.01. To obtain the root of these difference functions, we use a bisection algorithm. In order to further decrease the noise level, we compute the difference functions across pairs of simulations, each pair using an independent sequence of pathogens ${a_t}$ of length $L = 400$. The number of independent pairs of simulations used for each value of $\sigma_v$ and $\phi$ is $M \sim 10^{5}$.

\section*{Acknowledgements}
The study was supported by the European Research Council COG 724208
and ANR-19-CE45-0018 ``RESP-REP'' from the Agence Nationale de la Recherche and DFG grant CRC 1310
``Predictability in Evolution''.
The authors thank Natanael Spisak for pre-processing the raw data from \cite{Briney2019}, and for useful discussions and suggestions.

\bibliographystyle{pnas}

\onecolumngrid

\appendix

\renewcommand{\thefigure}{S\arabic{figure}}
\setcounter{figure}{0}

\section*{Supplementary information}

\section{Mean-field naive coverage}

Here we show how the infection cost function defined in the main text,
\beq\label{eq:infection}
I_t=\min\left[\phi, \left(\sum_{x\in P_{t-1}}c_{x,t}f(x,a_t)\right)^{-\alpha} \right],
\eeq
may be derived as the mean-field limit of a repertoire with memory and naive compartments.

In addition to the evolving memory repertoire $P_t$ already described in the main text, we define a naive repertoire made of random receptors $\mathcal{N}$, distributed uniformly with density $\rho$. Viruses may be recognized by either the memory or naive clonotypes. 
The naive coverage is defined as:
  \beq
  C_{\rm naive}(a_t)=\sum_{x\in\mathcal{N}} f(x,a_t),
  \eeq
  and the memory coverage as before:
    \beq
  C(a_t)=\sum_{x\in P_{t-1}} c_{x,t}f(x,a_t).
  \eeq
(In this convention, each naive clonotype has size one in arbitrary units.)

Depending on the values of these coverages, the system will choose to use either the naive repertoire, or an existing memory. In this decision, we factor in the fact that using the naive repertoire is more costly, which we account for using a prefactor $\beta<1$.
The cost is then defined as:
  \beq
  L_t=\max\left[\beta C_{\rm naive}(a_t),C(a_t)\right]^{-\alpha}.
  \eeq

We can simplify this expression in the limit where naive clonotypes are very numerous, but each offer weak coverage.
In the limit of high density of naive cells, $\rho\to \infty$, the coverage self-averages to its mean value:
  \beq
  C_{\rm naive}\approx \<C_{\rm naive}\>=\rho \int d^d x\, f(x,a_t)=\rho U_d(q) r_0^d,
  \eeq  
with
\beq
U_d(q)= \int d^d y\,e^{-\Vert y\Vert^q}=S_d \int r^{d-1}dr\, e^{-r^q},
\eeq
where we have done the change of variable $x=a_t+yr_0$, and where $S_d=2\pi^{d/2}/\Gamma(d/2)$ is the surface area of the unit sphere.

Taking the $\rho\to\infty$ and $\beta\to 0$ limits, while keeping $\beta\rho$ finite, corresponds to a dense naive repertoire but where each naive cell weakly covers the antigenic space. In this limit we recover the model of the main text
\beq
I_t=\min\left[\phi,C(a_t)^{-\alpha}\right],
\eeq
with $\phi=(\rho\beta U_d(q)r_0^d)^{-\alpha}$.

\section{Transition from monoclonal to naive phase at $\sigma_v = 0$}
Here we derive an expression for the phase boundary between the monoclonal and polyclonal phases in the limit $\sigma_v=0$, where the virus does not move.

In the special case where $\mu=0$, clonotypes cannot multiply. At each time step, a number $m_n$ of new clonotypes at created at $a_n={\rm const}$, distributed according to a Poisson law of mean $\bar m$. This number is added to existing clonotypes, of which a random fraction $\gamma$ survives. If the previous number of clonotypes, $M_n$, is Poisson distributed with mean $\bar M_n$, the number of surviving ones $M_n'$ is also Poisson distributed with mean $\gamma\bar M_n$ (since subsampling a Poisson-distributed number still gives a Poisson law). Then, the new number of clonotypes, $M_{n+1}=M_n'+m_n$, is also Poisson distributed, with the recurrence relation:
\beq
\bar M_{n+1}=\gamma \bar M_n+\bar m.
\eeq
At steady state, we have
\beq
\bar M_{n}\to\frac{\bar m}{1-\gamma}.
\eeq

Since all clonotypes are at $x=a_n$, the coverage is $C(a_n)=M$, so that the expected cost reads:
\beq
\mathcal{L} = \phi\exp\left(-\frac{\bar m}{1 - \gamma} \right)  + \sum_{M = 1}^{+ \infty} \frac{1}{M^\alpha} \exp\left(-\frac{\bar m}{1 - \gamma} \right) \frac{1}{m!}{\left(\frac{\bar m}{1 - \gamma}\right)}^M + \kappa \bar m.
\eeq
To find the transition from monoclonal to naive, $\bar m=0$, we need to find the value of $\phi$ for which $\partial\mathcal{L}/\partial \bar m$ changes sign at $\bar m=0$: if this derivative is positive, it is better to have $\bar m=0$ (since the function is convex); if it is negative, there is benefit to be gained by increasing $\bar m>0$. The condition:
\beq
\left.\frac{\partial\mathcal{L}}{\partial \bar m}\right|_{\bar m=0}=\kappa-\frac{\phi}{1-\gamma}+\frac{1}{1-\gamma}
\eeq
gives the transition point
\beq
\phi_c=1+\kappa(1-\gamma)
\eeq

For $\mu>0$, we redefine $M_n$ as the sum of all clonotype sizes, which is equal to the coverage, $C=M_n=\sum_{x\in P_n}c_{x,n}$. The recurrence relation is replaced by:
\beq
\bar M_{n+1}=\gamma (1+\mu) \bar M_n+\bar m.
\eeq
For $\gamma(1+\mu)>1$, this number explodes, so that $M$ is infinite, reducing the infection cost to 0 regardless of $\bar m$. The transition point is then
\beq
\phi_c=0.
\eeq

For $\gamma(1+\mu)<1$, $\bar M_n$ reaches a steady state value,
\beq
\bar M_n\to \frac{\bar m}{1-\gamma(1+\mu)}.
\eeq
Although $M_n$ is not strictly distributed according to a Poisson law, it is still a good approximation, so that we can repeat the same argument as with $\mu=0$,
\beq
\phi_c\approx 1+\kappa(1-\gamma(1+\mu)).
\eeq

\begin{figure*}
\begin{center}
\includegraphics{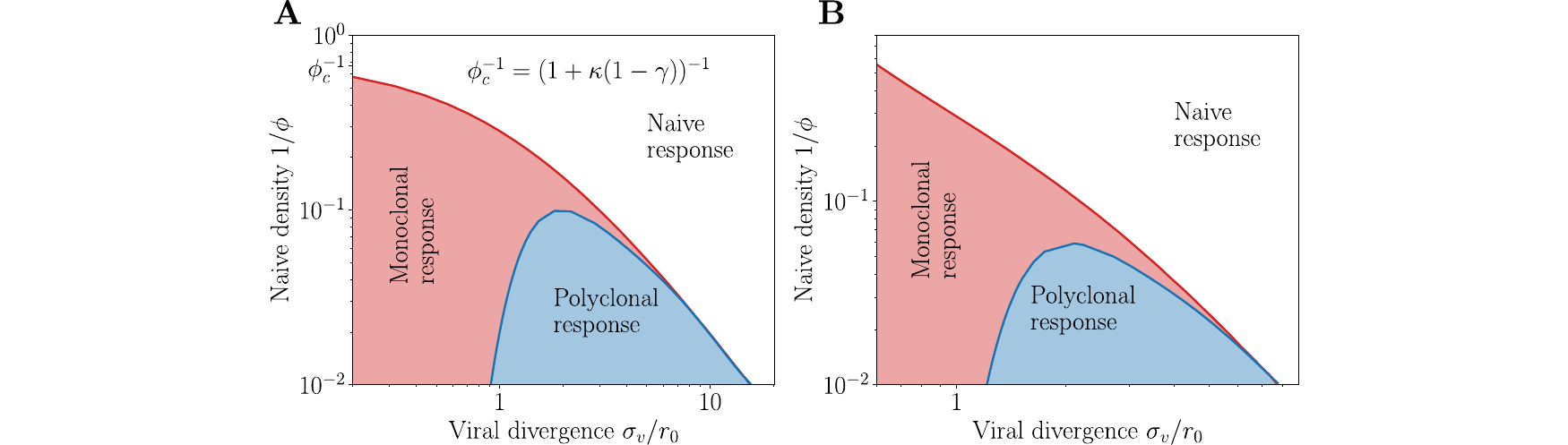}
\caption{{\bf Phase diagram for various parameters.} {\bf A.} Phase diagram for parameters $\mu = 0$, $\gamma = 0.85$ and $d = 2$. We observe the upper transition point $\phi_c = (1 + \kappa (1 - \gamma))^{-1}$ at $\sigma_v  = 0$. {\bf B.} Phase diagram for parameters $\mu = 0.5$, $\gamma = 0.85$ and $d = 3$. Since $\gamma(1 + \mu) > 1$ the transition point $\phi_c = \infty$. In both {\bf A} and {\bf C} we observe that the phase diagram retains the same shape. In this panel $\alpha = 1$, $q = 2$ and $\kappa = 0.5/(1 - \gamma)$.}
\label{fig:figS0}  
\end{center}
\end{figure*}

{\ch

\section{Analytical results in a solvable model}

\begin{figure*}[t!]
\begin{center}
\includegraphics{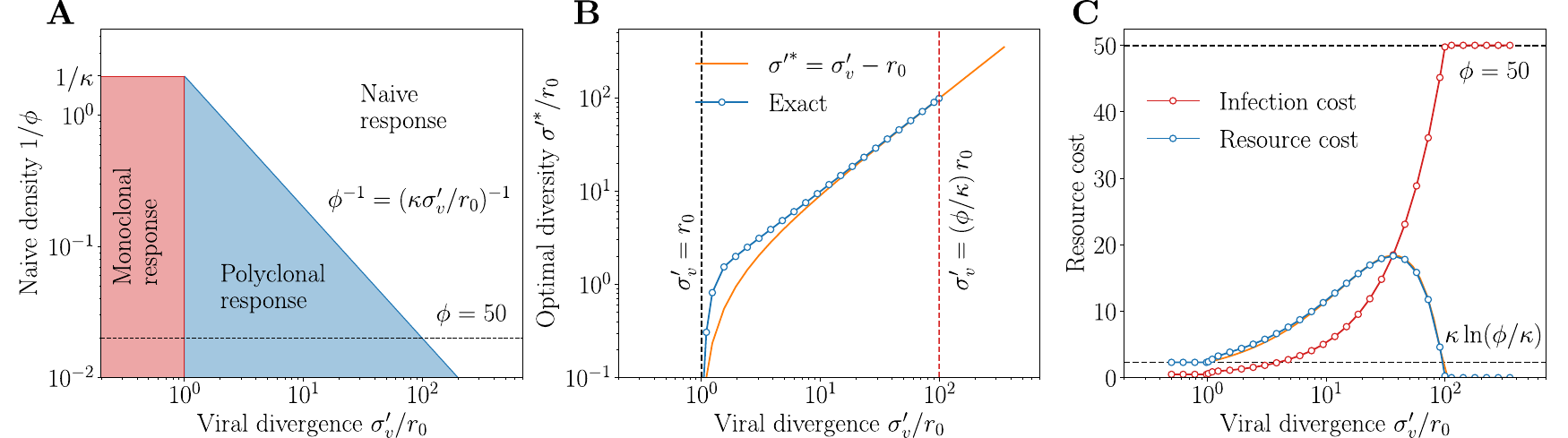}
\caption{{\bf Analytical solution of a tractable model. A.} Exact phase diagram in $d = 1$ for the simplified model ($q = \infty$, $\gamma=0$, and all-or-nothing infection cost). The boundary between monoclonal is given by $\sigma'_v = r_0$ and the boundary between polyclonal by $\phi^{-1} = (\kappa \sigma_v'/r_0)^{-d}$. {\bf B.}  Optimal memory diversity $\sigma'^*\approx \sigma_v'$ and {\bf C.} optimal infection and plasticity costs for as a function of $\sigma'_v$  for $\phi = 50$. $\sigma'$ and $\sigma'_v$ are rescaled versions of the diversity and divergence to match the variances of the original model.}
\label{fig:fig2}
\end{center}
\end{figure*}

\subsection{Model definition and main results}

To gain insight into the transitions observed in the phase diagram of Fig.~2, we can make a series of simplifications and approximations about the model that allow for analytical progress. We assume a step function for the cross-reactivity function $f(x,a)=1$ for $\Vert x-a\Vert\leq r_0$, and 0 otherwise, corresponding the limit $q= \infty$. Likewise, we assume a uniform distribution of viral antigenic mutations $a_{t+1}=a_t+\sigma_v' \eta'_{t+1}$, where $\eta'_t$ is a random point of the $d$-dimensional unit ball, with $\sigma_v'=\sigma_v\sqrt{1+2/d}$ (so that the variance is the same as in the Gaussian case), and similarly for memory diversification, with new clonotypes drawn from a uniform distribution is a ball of radius $\sigma'=\sigma\sqrt{1+2/d}$. The infection cost is approximated by an all-or-nothing function, with $I_t=0$ if there is any coverage $C(a_t)>0$, and $I_t=\phi$ if $C(a_t)=0$. We further assume $\gamma=0$: all clonotypes are discarded at each time step, so that memory may only be used once. 

In this simplified version of the model, the phase diagram and optimal parameters can be computed analytically. We first summarize the main results below. In the subsequent paragraphs, we provide detailed derivations in the case of arbitrary dimensions, and also provide additional exact formulas in the one-dimensional case.

One can show (see next paragraphs) that the transition from monoclonal to polyclonal response occurs exactly when the radius of the ball within which viral mutations occur reaches the cross-reactivity radius $r_0$:
\beq
\sigma_{v}'=r_0.
\eeq
Below this transition ($\sigma_v'<r_0$), the optimal strategy is to have no diversity at all and perfectly target the recognized antigen $a_t$, ${\sigma'}^*=0$, as any memory cell at $a_t$ will recognize the next infection. In this case the optimal mean number of memories $\bar m^*=\ln(\phi/\kappa)$ results in a trade off between the cost of new memories with the risk of not developing any memory at all by minimizing $\phi e^{-\bar m}+\kappa \bar m$. The transition from the monoclonal response to naive phases is then given by $\phi=\kappa$, where $\bar m^*=0$.

The polyclonal-to-naive transition may also be understood analytically. In the polyclonal response phase, the optimal strategy is, in either of the limits $\sigma'_v \gg r_0$ or $\bar m^* \ll 1$ (see next paragraphs):
\begin{align}
{\sigma'}^*&\approx \sigma_v'-r_0 \label{sigmastar}\\
\bar m^*&\approx \frac{\sigma_v'^d}{r_0^d}\ln \left(\frac{\phi}{\kappa}\frac{r_0^d}{\sigma_v'^d}\right). \label{mstar}
\end{align}
In particular this result becomes exact at the transition from polyclonal to naive, where $\bar m^*=0$. The transition is thus given by:
\beq
\phi^{-1} = \frac{r_0^d}{\kappa\sigma_v'^d}.
\eeq
The polyclonal response is outcompeted by the naive one when the density of naive cell ($\phi^{-1}$) becomes larger than the probability density of new strains falling within the cross-reactivity radius ($r_0^d/{\sigma_v'}^d$), rescaled by the memory cost coefficient $\kappa^{-1}$.

Fig.~\ref{fig:fig2} shows the resulting phase diagram, as well as the optimal diversity $\sigma'^*$ and predicted costs for a fixed $\phi$ and $d=1$. These predictions reproduce the main features of the full model, in particular the scaling of the immune diversity $\sigma$ with $\sigma_v$ (Fig.~\ref{fig:fig1}D vs. Fig.~\ref{fig:fig2}B) and the general shape of the optimal memory size $\bar m^*$ (Fig.~\ref{fig:fig1}E vs. Fig.~\ref{fig:fig2}C), which first increases as the virus becomes more divergent, to later drop to zero as memory becomes too costly to maintain and the system falls into the naive phase. 

  }

\subsection{General formulation}
Define $P_{\rm hit}(\sigma',r_0,r)$ as the probability that a random memory will recognize the next infection at distance $r$, i.e. the probability that a random point in the ball of radius $\sigma'$ and a point at distance $r$ from its center are at distance $\leq r_0$. The probability that none of $m$ clonotypes recognize the virus, where $m$ is drawn from a Poisson distribution of mean $\bar m$, reads:
\beq
P_{\rm miss}(\sigma',r_0,r)=\sum_m e^{-\bar m}\frac{\bar m^m}{m!}(1-P_{\rm hit}(\sigma',r_0,r))^m=e^{-\bar m P_{\rm hit}(\sigma',r_0,r)}.
\eeq
The best strategy maximizes this probability, averaged over the location of the next infection, over $\sigma'$:
\beq\label{eq:hatpmiss}
\bar P_{\rm miss}(\sigma',\bar m,r_0,\sigma_v')=\left\<e^{-\bar m P_{\rm hit}(\sigma',r_0,r)}\right\>_{\mathcal{B}(\sigma_v')}=\frac{1}{\sigma_v'^dV_d}\int_{0}^{\sigma_v'} S_d r^{d-1} dr\,e^{-\bar m P_{\rm hit}(\sigma',r_0,r)}
\eeq
where $\mathcal{B}_{\sigma_v'}$ is the ball of radius $\sigma_v'$,
$V_d=\pi^{d/2}/\Gamma(d/2+1)$ is the volume of a unit ball, and $S_d=2\pi^{d/2}/\Gamma(d/2)$ the area of the unit sphere.

Then the expected overall cost reads:
\beq\label{eq:overall}
\mathcal{L}= \phi \bar P_{\rm miss}(\sigma',\bar m,r_0,\sigma_v) + \kappa \bar m =\phi \left\<e^{-\bar m P_{\rm hit}(\sigma',r_0,r)}\right\>_{\mathcal{B}(\sigma_v')}
 + \kappa \bar m.
\eeq

\subsection{Exact location of the phase transitions, and approximate solution in the polyclonal phase}
The location of the optimal $\sigma'$ may be rigorously bounded from above and below.
If $\sigma'<\sigma'_v-r_0$, then only part of the future positions of the virus are covered, so increasing $\sigma'$ can bring no harm. Likewise, for $\sigma'>\sigma'_v+r_0$, memory covers parts of the antigenic space that have no chance of harboring the next virus, so that decreasing $\sigma'$ is also always advantageous. Thus, the optimum $\sigma'^*$ must satisfy:
\beq\label{ineq}
\sigma_v'-r_0\leq \sigma'^*\leq \sigma_v'+r_0.
\eeq

As already argued in the main text, when $\sigma_v'<r_0$, there is clearly no benefit to having $\sigma'>0$, so the optimum is reached at $\sigma'=0$. \eqref{ineq} further shows that if $\sigma_v'>r_0$, then $\sigma'^*>0$, so that a polyclonal phase is optimal. As a consequence, the transition from the monoclonal to polyclonal phases happens exactly at
\beq
\textrm{Monoclonal to polyclonal:}\quad\sigma_v'=r_0.
\eeq

In the monoclonal phase, memory always recognizes the next virus. The only risk of paying $\phi$ is when no memory is created, which happens with probability $e^{-\bar m}$, so that the cost reads:
\beq
\mathcal{L}=\phi e^{-\bar m}+\kappa \bar m.
\eeq
The optimal $\bar m^*=\ln(\phi/\kappa)$ cancels at the monoclonal-to-naive transition:
\beq
\textrm{Monoclonal to naive:}\quad\phi =\kappa.
\eeq

In the polyclonal phase, we could not find a general analytical solution, but there are two limits in which the solution may be calculated. The first limit is when $\sigma_v\gg r_0$. In that case, \eqref{ineq} implies $\sigma'^*\approx \sigma_v'$, and
\beq
P_{\rm hit}(\sigma'^*,r_0,r)\approx \frac{r_0^d}{{\sigma}_v'^d},
\eeq
which doesn't depend on $r$. Then, minimizing
\beq
\mathcal{L}\approx\phi e^{-\bar m P_{\rm hit}}
 + \kappa \bar m
\eeq
with respect to $\bar m$ yields:
\beq
\bar m^*=\frac{1}{P_{\rm hit}}\ln\left(\frac{\phi}{\kappa}P_{\rm hit}\right)=\frac{\sigma_v'^d}{r_0^d}\ln \left(\frac{\phi}{\kappa}\frac{r_0^d}{\sigma_v'^d}\right).
\eeq

The second limit in which things simplify is for small $\bar m$. Then the exponential in $\eqref{eq:hatpmiss}$ may be expanded at first order, yielding:
\beq
\mathcal{L}=\phi\left(1-\bar m\<P_{\rm hit}(\sigma',r_0,r)\>_{\mathcal{B}(\sigma_v')}\right)+\kappa \bar m,
\eeq
where $\<\cdot\>_r$ is the mean of over the ball of radius $\sigma_v'$.
Minimizing with respect to $\sigma'$ is equivalent to maximizing $\<P_{\rm hit}(\sigma',r_0,r)\>_r$, which is the probability that a random point in the ball of radius $\sigma'$ and a random point in the ball of radius $\sigma_v'$ are separated by less than $r_0$. This probability is maximized for any $\sigma'\leq \sigma_v-r_0$, where it is equal to $(r_0/\sigma_v')^d$. Increasing $\sigma'$ beyond $\sigma_v-r_0$ can only lower the probability of recognition. Thus:
\beq
\min_{\sigma'}\mathcal{L}=\phi+\bar m\left(\kappa-\phi \frac{r_0^d}{{\sigma}_v'^d}\right).
\eeq
This gives us the condition for the transition from polyclonal to naive, where $\bar m^*=0$. This happens when
\beq
\textrm{Polyclonal to naive:}\quad\phi= \kappa \frac{{\sigma}_v'^d}{r_0^d}.
\eeq
This condition  gives us an exact expression for the location of the transition.

\subsection{Exact solution in dimension 1}
For $d=1$, the cost $\mathcal{L}$ in \eqref{eq:overall} may be calculated analytically, by using exact expressions of $P_{\rm hit}(\sigma',r_0,r)$. When $\sigma_v'\leq r_0$, the optimal $\sigma'$ is zero as explained in the main text. When $\sigma_v'> r_0$, we distinguish two cases: $r_0<\sigma_v'\leq 2r_0$, and $\sigma_v'>2r_0$.

\paragraph{Case $r_0<\sigma_v'\leq 2r_0$.}
Since we know that the optimal $\sigma'$ is between $\sigma_v'-r_0$ and $\sigma_v'+r_0$, we focus on that range. Then there are two subcases for $\sigma$'.

If $\sigma_v'-r_0<\sigma'\leq r_0$, there are two contributions to the integral of $\bar P_{\rm miss}$ over the position of the virus $r$. Either $r\leq r_0-\sigma'$, then all memories recognize the virus, $P_{\rm hit}=1$; or $r_0-\sigma'<r<\sigma_v'<r_0+\sigma'$, in which case the recognition probability is given by the normalized intersection of two balls at distance $r$ of radii $\sigma'$ and $r_0$,
\beq\label{eq:phit1}
P_{\rm hit}=\frac{\sigma'+r_0-r}{2\sigma'}.
\eeq
Thus we obtain doing the integral over $r$ in \eqref{eq:hatpmiss}:
\beq\label{eq:P1}
\bar P_{\rm miss}(\sigma',\bar m,r_0,\sigma_v')= \frac{1}{\sigma'_v} \left[ (r_0 - \sigma') e^{-\bar m} + \int_{r_0 - \sigma'}^{\sigma'_v} \exp \left( -\bar m \frac{\sigma'+r_0-r}{2\sigma'} \right) dr \right]  \text{ if } \sigma_v'-r_0<\sigma'\leq r_0.
\eeq

If $r_0<\sigma'\leq \sigma_v'+r_0$, there are also two contributions. Either $r\leq \sigma'-r_0$, in which case there is no boundary effect, and the recognition probability is just $P_{\rm hit}=r_0/\sigma_v'$; or $\sigma'-r_0<r\leq \sigma_v'\leq \sigma'+r_0$, in which case we have again \eqref{eq:phit1}. Performing the integration in \eqref{eq:hatpmiss} we obtain:
\beq\label{eq:P2}
\bar P_{\rm miss}(\sigma',\bar m,r_0,\sigma_v') = \frac{1}{\sigma'_v} \left[ \exp \left( - \bar m \frac{r_0}{\sigma'} \right) (\sigma' - r_0) + \int_{\sigma' - r_0}^{\sigma'_v} \exp \left( -\bar m \frac{\sigma' + r_0-r}{2 \sigma'} \right) dr \right] \text{ if } r_0<\sigma'\leq \sigma_v'+r_0.
\eeq
Numerical analysis shows that \eqref{eq:P1} admits a minimum as a function of $\sigma'$ in its interval of validity, $\sigma_v'-r_0<\sigma'^*\leq r_0$, while \eqref{eq:P2} is always increasing.

\paragraph{Case $\sigma_v'>2r_0$.}
In this case, there is only a single subcase in the range of interest $\sigma_v'-r_0$ and $\sigma_v'+r_0$. This case is the same as the previous one considered, and the result is given by the same formula \eqref{eq:P2}. However, for $\sigma_v'>2r_0$, this expression now admits a minimum $\sigma_v'-r_0<\sigma'^*\leq \sigma_v'+r_0$.

We recover that in both cases (a and b), in the limit $\bar m\to 0$, this minimum is reached at $\sigma_v'-r_0$.

\section{Population dynamics in sequential immunization}

\subsection{Clonotype growth and decay as a first-passage problem}

We now want to study clonotype proliferation induced by a recall response. We focus on the limit of small mutation rates $\sigma_v \ll r_0$. Within this regime, the system is in the monoclonal phase with $\sigma^* = 0$. We can therefore focus on the case of a single clonotype at position $x = 0$ on the phenotypic space, and ask how successive challenges will modify its size.
(Different initial conditions will only change the prefactor in front of the exponential modes in the distribution of first passage times, so the large time behavior of this probability distribution will be the same as discussed below.)

The clonotype has an initial size $c = 1$, and the virus drifts away from $x=0$ with viral divergence $\sigma_v$. In the general model, cells have probability $\gamma$ to survive from one challenge to the other. Proliferation is taken to be proportional to the cross reactivity radius, $\mu e^{-(r/r_0)^q}$. The population dynamics is thus given by the approximate recursion:
\beq
n_{t+1} \approx n_{t} \gamma \left[ 1 + \mu e^{-(r/r_0)^q} \right],
\eeq
where we have neglected birth-death noise.
We can further simplify this equation to $n_{t+1} = \gamma (1 + \mu \Theta(r - r^*) ) n_t$, where $r^*=r_0\ln(\gamma\mu/(1-\gamma))^{1/q}$ is defined as the radius at which the net fold-change factor crosses $1$, i.e. when birth is exactly compensated by death. This means that, as long as the virus is within distance $r^*$, the clonotype grows with fold-change factor $\sim \Gamma$. As soon as it reaches $r^*$, and neglecting possible returns below $r^*$ (which happen with probability 1 for $d\leq 2$, but with a frequency that does not affect the overall decay), it will decay with fold-change factor $\sim \gamma$. The problem is thus reduced to determining the first-passage time of the viral antigenic location at radius $r^*$.

We use a continuous approximation corresponding to a slowly evolving strain, $\sigma_v\ll r_0$:
\beq
a(0)=0,\qquad da=\frac{\sigma_v}{\sqrt{d}} dW,
\eeq
where $W$ is a Wiener process. The radius, given by $r(t)=|a(t)|$,
{\ch 
behaves on average as:
\beq
\<r(t)^2\>=t\sigma_v^2.
\eeq
The time it takes for $r(t)$ to reach $r^*$, denoted by $t^*$, is approximately given by $\<t^*\>\sim ({r^*}/{\sigma_v})^2$.

The tail of the distribution for this first-passage time is dominated by rare events when the virus mutates less than expected between infections, leading to larger episodes of growth. We will show in the next two sections that the distribution of these exceptionally long $t^*$ has an exponential tail:
\beq\label{eq:longtstar}
P(t^*>t)\sim e^{-t/t_s},\quad t_s\sim \<t^*\> \sim \frac{{r^*}^2}{\sigma_v^2}.
\eeq
This translates into a power-law tail for the peak clonotype abundance,
\beq\label{eq:beta}
p(n^*) \sim \frac{1}{{n^*}^{1+\beta}},\quad\textrm{with } \beta\sim \frac{\sigma_v^2}{{r^*}^2 \ln \Gamma}.
\eeq
The same scaling holds for the distribution of all abundances, since the peak determines the rest of the trajectory.

Within the same simplified picture, the lifetime $t_l$ of a clonotype is the sum of the time it takes to reach the peak, $t^*$, and the decay time until extinction, which is approximately $\ln(n^*)/\ln(1/\gamma)$:
\beq
t_l=t^*+\frac{\ln(n^*)}{\ln(1/\gamma)}=\left(1+\frac{\ln\Gamma}{\ln(1/\gamma)}\right)t^*.
\eeq
Thus, $t_l$ is proportional to $t^*$, and therefore also exponentially distributed:
\beq\label{eq:lambda}
p(t_l)\sim e^{-\lambda t_l},\quad \lambda\sim\frac{\sigma_v^2}{{r^*}^2}{\left(1+\frac{\ln\Gamma}{\ln(1/\gamma)}\right)}^{-1}.
\eeq

Next we derive in detail the distributions of the first passage time of $r(t)$ to $r^*$ to obtain Eq.~\ref{eq:longtstar}.
}

\subsection{First passage time in $d = 1$}

The distribution of first passage time, $p(t)$, can be computed solving diffusion with a box of size $2r^*$ in $d=1$ \cite{Redner2001}:
\beq
p(t)=\sum\limits_{n = 0}^{+\infty} \frac{(2n+1)\sigma_v^2\pi}{2{r^*}^2}(-1)^n \exp \left(-\frac{(2n+1)^2\pi^2 \sigma_v^2t}{8{r^*}^2} \right)
\eeq
The dominant term ($n=0$) at long times gives an exponential decay:
\beq
p(t) \approx \frac{\sigma_v^2\pi}{2{r^*}^2} \exp \bigg( -\frac{\pi^2 \sigma_v^2 t}{8{r^*}^2} \bigg).
\eeq

\subsection{First passage in higher dimensions}
We define $f(r,t)$ as the probability density that the virus has not yet reached $r^*$ at time $t$, and is at radius $r$.

This probability density is solution to the diffusion equation with spherical symmetry and absorbing boundary conditions in arbitrary dimension $d > 1$:
\beq\label{eq:diff}
\frac{\partial f}{\partial t} = \frac{\sigma_v^2}{2d} \left[ \frac{\partial^2 f}{\partial r^2} + \frac{d - 1}{r} \frac{\partial f}{\partial r} \right],\qquad f(r^*, t) = 0.
\eeq

Assuming separation of variables, $f(r, t) = T(t) R(r)$, we have:
\beq
\frac{T'(t)}{T(t)}=\frac{\sigma_v^2}{2d}\frac{R''(r)+\frac{d-1}{r}R'(r)}{R(r)}\equiv-\frac{\sigma_v^2}{2d{r^*}^2}\lambda,
\eeq
where $\lambda$ is to be determined later.
This implies $T(t) = Ce^{-\lambda \frac{\sigma_v^2}{2d{r^*}^2} t}$ where $C$ is a constant. The radial part $R(r)$ is solution to:
\beq
R''(r)+\frac{d-1}{r}R'(r) = - \frac{\lambda}{{r^*}^2} R(r).
\eeq
For $d = 1$ this equation reduces to a harmonic equation and we recover the above solution in 1D. Using the change of variable $R(r) = r^{1 - d/2} g(r)$ we derive the following equation:
\beq
r^2g''(r) + rg'(r) + \bigg(\lambda \frac{r^2}{{r^*}^2} - \bigg(\frac{d}{2} - 1\bigg)^2\bigg)g(r) = 0.
\eeq
Changing the variable $x = \sqrt{\lambda}r/r^*$, the function $\tilde g(x)=g(xr^*/\sqrt{\lambda})$ is solution to the Bessel differential equation of order $d/2 - 1$.  It can therefore be written as a superposition of a Bessel function of the first kind and a Bessel function of the second kind, both of order $d/2 - 1$. The Bessel function of the  second kind having a singularity at $x = 0$, our solution is only given by the Bessel function of the first kind $g(xr^*/\sqrt{\lambda}) = B J_{d/2 - 1} (x)$. The radial function $R$ now reads:
\beq
R(r) = B r^{1 - d/2} J_{d/2 - 1} (\sqrt{\lambda} r/r^*).
\eeq
The absorbing boundary condition at $r  = r^*$ gives us the condition $J_{d/2 - 1}(\sqrt{\lambda}) = 0 $, which has an infinite number of solutions $j_{0, d/2 - 1}, ..., j_{n, d/2 - 1}, ...$, so that $\lambda$ can take values
\beq
\lambda_n = j_{n, d/2 - 1}^2.
\eeq
The general solution to \eqref{eq:diff} is given as a linear combination of all possible modes, with coefficients $C_n$ determined from boundary conditions and the Dirac delta initial condition, $f(r,0) = \delta(r)$:
\beq
f(r,t) = \sum\limits_{n = 0}^{+\infty} C_{n} r^{1 - d/2} J_{d/2 - 1} \left(\frac{j_{n, d/2 - 1}}{r^*} r\right) \exp \left( - \frac{j_{n, d/2 - 1}^2}{2d{r^*}^2} \sigma_v^2 t \right).
\eeq
The distribution of first passage times asymptotically follows the largest mode of this series, $n=0$, so that:
\beq
p(t) \sim \exp \bigg(-\frac{j_{0, d/2 - 1}^2}{2d{r^*}^2} \sigma_v^2 t\bigg).
\eeq
For instance for $d=2,3,4$ we have $j_{0,0} \approx 2.40483$, $j_{0,1/2} = \pi$, $j_{0, 1} \approx 3.83171$.

\begin{figure*}
\begin{center}
\includegraphics{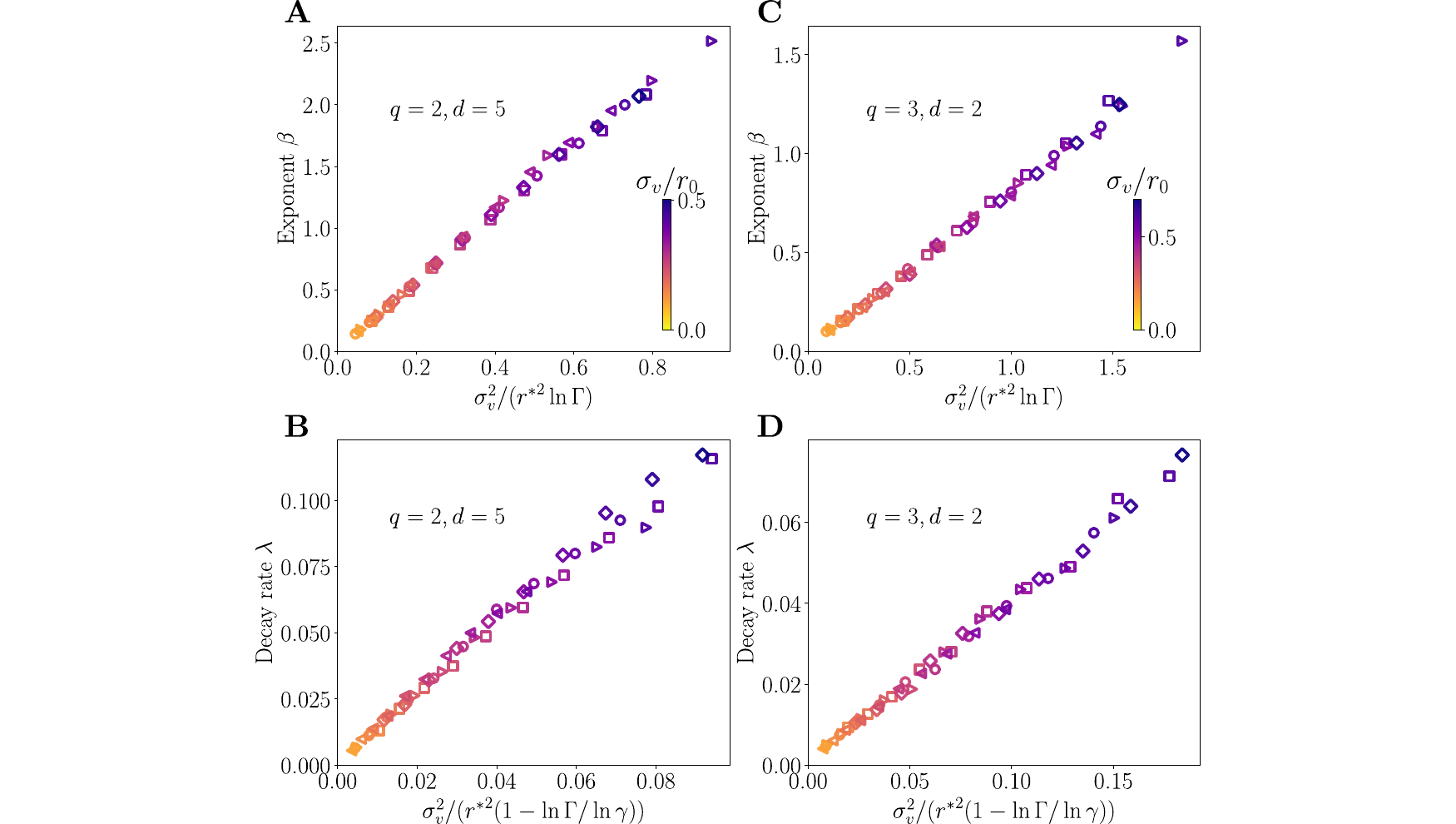}
\caption{{\bf Scaling relations for various parameters.} {\bf A-B.} Power law exponent $\beta$ and lifetime decay rate $\lambda$ in dimension $d = 5$ with a Gaussian cross-reactivity kernel with $q = 2$. {\bf C-D.} Power law exponent $\beta$ and lifetime decay rate $\lambda$ in dimension $d = 2$ with a cross-reactivity kernel with $q = 3$. The different parameters used are $(\gamma = 0.82, \mu = 0.65)$  i.e. $\Gamma = 1.353$ (diamonds), $(\gamma = 0.8, \mu = 0.62)$ i.e. $\Gamma = 1.296$ (squares), $(\gamma = 0.85, \mu = 0.5)$ i.e. $\Gamma = 1.275$ (circles), $(\gamma = 0.87, \mu = 0.4)$ i.e. $\Gamma = 1.21$ (triangles $>$), $(\gamma = 0.9, \mu = 0.35)$ ie. $\Gamma = 1.21$ (triangles $<$). The strategy is optimized with $\phi = 100$, $\kappa = 0.5/(1-\gamma)$. We used $\alpha = 1$ throughout.}
\label{fig:figS1}  
\end{center}
\end{figure*}

\begin{figure*}[t!]
\begin{center}
\includegraphics{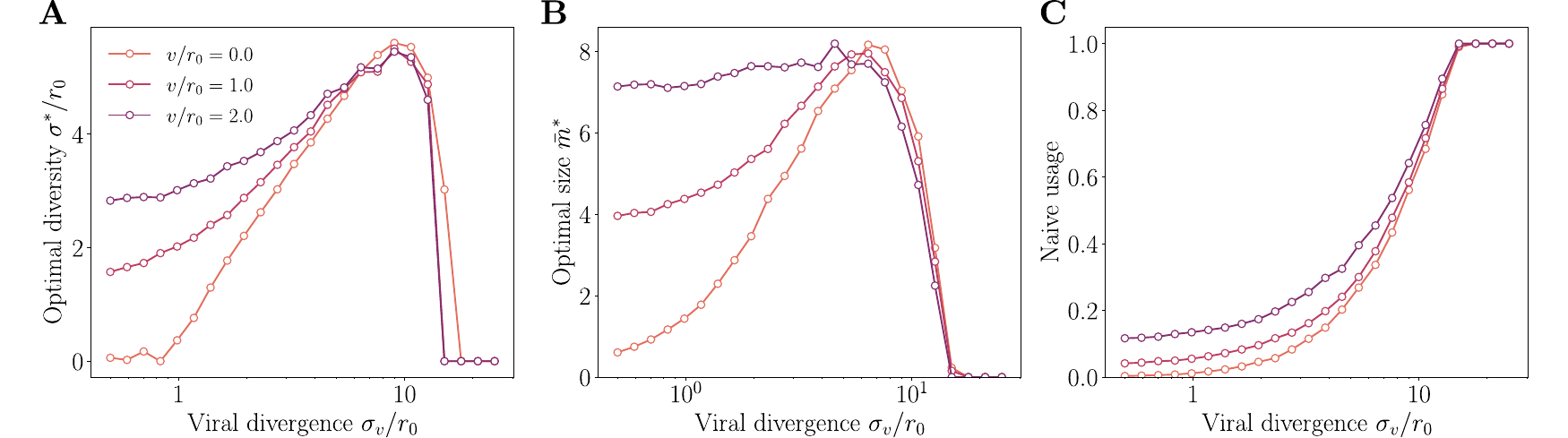}
\caption{{\bf Optimal strategy in presence of drift. A.} Optimal diversity {\bf B.} size and {\bf C.} frequency of naive usage in response to an immunization challenge for different strain drift $v/r_0$. The strain follows a random walk with drift $v$ in a fixed direction $e_0$: $a_{t+1} = a_t + v e_0 + \sigma_v \eta_t$. Parameters values: $\mu = 0.5$, $\gamma = 0.85$, $\kappa = 3.3$, $\alpha = 1$, $q = 2$, $d = 2$.}
\label{fig:figS2}
\end{center}
\end{figure*}

\end{document}